\documentclass[aps,prd,onecolumn,groupedaddress,floatfix,longbibliography,superscriptaddress]{revtex4-2}

\usepackage{amsmath}
\usepackage{amssymb}
\usepackage{amsfonts}
\usepackage{bbold}
\usepackage{bm}
\usepackage{cancel}
\usepackage{times,float}
\usepackage{graphicx}
\usepackage[usenames,dvipsnames,svgnames]{xcolor}
\usepackage{hyperref}
\hypersetup{colorlinks=true, linkcolor=Blue, citecolor=Blue,urlcolor=Blue}
\usepackage{multirow}
\usepackage{ulem}
\usepackage{color,epsfig}
\usepackage{slashed}
\usepackage{feynmf}
\usepackage{array}
\usepackage{physics}
\usepackage{subcaption}
\usepackage{soul}

\begin{document}

\title{Magnetic-field-activated transport from band geometry in gapped nodal-line semimetals}

\author{L. Medel Onofre}
\email{leonardo.medel@correo.nucleares.unam.mx}
\address{Instituto de Ciencias Nucleares, Universidad Nacional Aut\'{o}noma de M\'{e}xico, 04510 Ciudad de M\'{e}xico, M\'{e}xico}

\author{L. E. Sosa-Arias}
\email{sosa.luis@ciencias.unam.mx}
\affiliation{Instituto de Ciencias Nucleares, Universidad Nacional Aut\'{o}noma de M\'{e}xico, 04510 Ciudad de M\'{e}xico, M\'{e}xico}

\author{A. Mart\'{i}n-Ruiz}
\email{alberto.martin@nucleares.unam.mx}
\address{Instituto de Ciencias Nucleares, Universidad Nacional Aut\'{o}noma de M\'{e}xico, 04510 Ciudad de M\'{e}xico, M\'{e}xico}

\begin{abstract}
We develop a nonperturbative semiclassical theory of magnetotransport in gapped nodal-line semimetals, retaining the full magnetic-field dependence of Berry-curvature and orbital-magnetic-moment corrections. Starting from a minimal two-band model, we derive exact expressions for both the intrinsic Hall response and the dissipative Fermi-surface conductivity, valid to all orders in the magnetic field within the semiclassical regime. We show that the orbital magnetic moment reconstructs the geometrically active Fermi surface, breaking the azimuthal cancellation imposed by the nodal-line geometry and thereby activating an intrinsic Hall current that is absent at zero magnetic field. The Hall response exhibits a pronounced nonmonotonic dependence on the chemical potential and a strongly nonlinear magnetic-field evolution, reflecting the redistribution of geometrically active states around the nodal ring. We further demonstrate that the Fermi-surface conductivities recover the conventional Drude behavior in the weak-field limit, while acquiring quadratic and ultimately nonperturbative magnetic-field corrections together with a field-induced transport anisotropy that gives rise to a measurable planar Hall effect at intermediate fields. Both the intrinsic and dissipative responses are strongly enhanced when the Fermi level lies close to the nodal ring, where Berry curvature and orbital magnetic moment are largest. Our results identify magnetic-field-activated geometric transport as a characteristic signature of nodal-line semimetals and provide a unified framework for describing magnetotransport beyond perturbative magnetic-field expansions, together with experimentally accessible signatures of orbital-magnetic-moment physics.
\end{abstract}

\maketitle

\section{Introduction}

Topological semimetals have emerged as an important platform for exploring the interplay between electronic band topology and transport phenomena~\cite{RevModPhys.82.3045, RevModPhys.83.1057, RevModPhys.90.015001, 10.1146/annurev031016-025458}. In these systems, conduction and valence bands touch near the Fermi energy, giving rise to low-energy quasiparticles whose dynamics are governed not only by the band dispersion but also by geometric quantities such as the Berry curvature and the orbital magnetic moment~\cite{RevModPhys.82.1959, PhysRevLett.97.026603, 10.1142/S0217984906010573}. These geometric properties generate a broad range of unconventional responses, including anomalous Hall effects, planar Hall effects, nonlinear transport, magneto-optical phenomena, and magnetic-field-induced electronic currents~\cite{RevModPhys.82.1539, PhysRevLett.112.166601, PhysRevB.91.214405}, establishing topological semimetals as an ideal setting for investigating geometry-driven transport beyond conventional band theory.

Among topological semimetals, nodal-line semimetals occupy a distinctive position. Unlike Weyl or Dirac semimetals, where band crossings occur at isolated points in momentum space, nodal-line semimetals host extended band degeneracies forming closed loops protected by crystal symmetries or combined inversion and time-reversal symmetry~\cite{PhysRevB.84.235126, PhysRevLett.115.036806, Fang2016TNLSM, Bian2016}. When spin-orbit coupling opens a small gap, the Berry curvature and orbital magnetic moment become strongly localized around the nodal ring while preserving their characteristic toroidal distribution in momentum space~\cite{PhysRevB.97.161113, Yang31122022}. The resulting Fermi surface geometry differs qualitatively from that of point-node semimetals, providing a natural platform in which geometric effects can be enhanced and manipulated through external magnetic fields. {Beyond their unusual bulk topology, nodal-line semimetals also support drumhead surface states and exhibit a remarkably rich transport phenomenology, including interface scattering, resonant transmission, and electron refraction, as demonstrated in recent theoretical studies~\cite{Rudi2024}.}

The influence of band geometry on transport is naturally described within the semiclassical wave-packet formalism~\cite{RevModPhys.82.1959, PhysRevLett.97.026603, 10.1142/S0217984906010573}. Within this approach, Berry curvature modifies the equations of motion through anomalous velocity terms and corrections to the phase-space measure, while the orbital magnetic moment introduces a magnetic-field-dependent correction to the quasiparticle spectrum~\cite{PhysRevLett.112.166601, PhysRevB.91.214405}. Together, these ingredients determine the electronic dynamics in external electromagnetic fields and provide the foundation for describing intrinsic transport phenomena in topological materials. In gapped nodal-line semimetals, where both the Berry curvature and orbital magnetic moment are concentrated around the nodal loop, the transport response becomes particularly sensitive to the geometry of the Fermi surface and its evolution under applied magnetic fields. {Recent semiclassical studies have further demonstrated that these geometric quantities also govern thermoelectric, magnetothermal, and Hall transport in topological semimetals, emphasizing the central role played by the orbital magnetic moment in magnetic-field-driven responses~\cite{Haidar2025}.}

Most theoretical studies of magnetotransport in nodal-line semimetals have focused on the weak-field regime, where transport coefficients are obtained through perturbative expansions in powers of the magnetic field~\cite{PhysRevLett.112.166601, PhysRevB.91.214405, PhysRevB.98.155125, Flores-Calderon_2023, Ipsita_Gapped_NLS}. Although this approach successfully captures the leading corrections to the conductivity tensor, it necessarily neglects the full nonlinear dependence of the semiclassical dynamics on the magnetic field. {Recent studies have further shown that the interplay between Berry curvature and the orbital magnetic moment gives rise to anisotropic linear magnetoelectric transport within this perturbative framework~\cite{Ipsita_Gapped_NLS}.} {Complementary theoretical approaches based on nonequilibrium Green-function and Landauer-B\"uttiker formalisms have also been employed to investigate magnetoelectric transport and finite-size effects in nodal-line systems, providing an alternative description of quantum transport in mesoscopic regimes~\cite{Cheng2024}.} In particular, the orbital magnetic moment modifies both the quasiparticle dispersion and the topology of the Fermi surface, thereby changing the phase space available for transport. As the magnetic field increases, these geometric corrections become intrinsically nonperturbative and cannot be faithfully represented by any finite-order expansion.

An especially intriguing consequence of the orbital magnetic moment is its ability to activate Hall-like responses that are otherwise forbidden by symmetry. In nodal-line semimetals, the Berry curvature circulates around the nodal loop, leading to an exact cancellation of transverse contributions after integration over all azimuthal directions. An external magnetic field breaks this balance indirectly through the orbital magnetic moment, which reconstructs the geometrically active regions of the Fermi surface and redistributes the occupation of the electronic states that contribute most efficiently to transport. Consequently, finite Hall currents emerge without requiring any additional symmetry breaking beyond the applied magnetic field. Since the same mechanism simultaneously generates an anisotropic longitudinal conductivity, it naturally gives rise to a planar Hall effect whose origin is entirely geometric~\cite{PhysRevLett.119.176804, PhysRevB.96.041110, PhysRevB.108.155132, Medel_Sci_Reports}. {Very recently, planar Hall transport in gapped nodal-line semimetals has also been investigated within the linear-response regime, further highlighting the role of Berry curvature and the orbital magnetic moment in generating anisotropic transport in these systems~\cite{Ipsita_Gapped_NLS}.}

In this work we develop a fully nonperturbative semiclassical theory of magnetotransport in gapped nodal-line semimetals. While previous studies have successfully described intrinsic Hall, planar Hall, and anisotropic magnetoelectric responses within perturbative weak-field expansions~\cite{Ipsita_Gapped_NLS, PhysRevB.108.085120}, these approaches necessarily rely on low-order expansions of the semiclassical dynamics and therefore cannot capture the complete nonlinear interplay between Berry curvature, the orbital magnetic moment, and the magnetic-field-induced reconstruction of the Fermi surface. In contrast, our formulation retains the full magnetic-field dependence of the phase-space factor, generalized velocity, and orbital-magnetic-moment-corrected energy spectrum without resorting to any weak-field approximation. Starting from a minimal two-band Hamiltonian, we derive exact analytical expressions for both the intrinsic Hall response and the dissipative Fermi-surface conductivity that remain valid to all orders in the magnetic field within the semiclassical regime. Furthermore, we demonstrate that the field-induced transport anisotropy naturally gives rise to a planar Hall effect, providing a direct and experimentally accessible signature of the geometric reconstruction of the Fermi surface. The resulting formalism establishes a unified framework for describing magnetic-field-activated geometric transport beyond the conventional weak-field approximation.

Our results demonstrate that the magnetic field plays a dual role. Besides modifying the quasiparticle dynamics, it continuously reconstructs the effective Fermi surface, redistributing the geometrically active electronic states responsible for transport. This reconstruction produces magnetic-field-activated Hall currents together with strongly nonlinear conductivity corrections whose magnitude depends sensitively on the position of the chemical potential relative to the nodal ring. We further show that the resulting transport anisotropy naturally gives rise to a planar Hall effect, providing a direct and experimentally accessible manifestation of the magnetic-field-driven geometric reconstruction of the Fermi surface. Our work therefore establishes a unified nonperturbative description of geometric magnetotransport in nodal-line semimetals beyond the conventional weak-field approximation.

The remainder of this paper is organized as follows. In Sec. \ref{Boltzmann_Section} we introduce the semiclassical Boltzmann formalism and derive the decomposition of the electric current into intrinsic and Fermi-surface contributions. In Sec. \ref{Model_NLSM} we present the nodal-line semimetal model and discuss its geometric properties, including the Berry curvature and orbital magnetic moment. In Sec. \ref{Current_Section} we evaluate the intrinsic Hall response and the dissipative magnetotransport coefficients, obtaining exact expressions valid to all orders in the magnetic field and comparing them with their weak-field limits. Finally, in Sec. \ref{conclusion} we summarize our main results and discuss their physical implications.

\section{Semiclassical Boltzmann description}
\label{Boltzmann_Section}

We analyze transport in nodal-line semimetals by means of a semiclassical approach in which charge carriers are described as wave packets evolving under external electromagnetic fields $(\mathbf{E},\mathbf{B})$. In contrast to systems with point-like band crossings, the low-energy structure is instead organized along extended nodal loops in momentum space. Consequently, geometric quantities such as the Berry curvature and the orbital magnetic moment are not localized but rather distributed along these loops, generally exhibiting a pronounced anisotropy. {This extended distribution has recently been shown to underlie a variety of Hall, thermoelectric, and magnetothermal transport phenomena in topological semimetals~\cite{Haidar2025}.}

The dynamics of a wave packet in band $n$, centered at $(\mathbf{r},\mathbf{k})$, is governed by the semiclassical equations of motion \cite{RevModPhys.82.1959}
\begin{align}
\dot{\mathbf{r}} _{n} &= D _{n\mathbf{k}} \left[ \mathbf{v} _{n\mathbf{k}} + \frac{e}{\hbar} \, \mathbf{E} \times \boldsymbol{\Omega} _{n\mathbf{k}} + \frac{e}{\hbar} \, (\mathbf{v} _{n\mathbf{k}} \cdot \boldsymbol{\Omega} _{n\mathbf{k}}) \, \mathbf{B} \right] , \label{rdot_decoupled} \\ 
\hbar \dot{\mathbf{k}} _{n} &= D _{n\mathbf{k}} \left[ - e \mathbf{E} - e \, \mathbf{v} _{n\mathbf{k}} \times \mathbf{B} - \frac{e^{2}}{\hbar} \, (\mathbf{E} \cdot  \mathbf{B}) \, \boldsymbol{\Omega} _{n\mathbf{k}} \right] , \label{kdot_decoupled}
\end{align}
where $\mathbf{v} _{n\mathbf{k}} = \hbar ^{-1} \nabla _{\mathbf{k}} \, \mathcal{E} _{n \mathbf{k}}$ is the band velocity and 
\begin{equation}
D _{n\mathbf{k}}
=
\left(
1+
\frac{e}{\hbar}\,
\mathbf{B}\cdot \boldsymbol{\Omega}_{n\mathbf{k}}
\right)^{-1} 
\end{equation}
is the Berry-curvature correction to the phase-space measure \cite{PhysRevLett.97.026603, 10.1142/S0217984906010573}. The Berry curvature is defined as
\begin{equation}
\boldsymbol{\Omega} _{n\mathbf{k}}
=
i
\bra{\nabla_{\mathbf{k}} u_{n \mathbf{k}}}
\times
\ket{\nabla_{\mathbf{k}} u_{n \mathbf{k}}}, \label{Berry_curvature}
\end{equation}
and the band energy includes a correction from the orbital magnetic moment $\mathbf{m} _{n \mathbf{k}}$, i.e. $\mathcal{E} _{n \mathbf{k}} = E _{n \mathbf{k}}  - \mathbf{m} _{n \mathbf{k}} \cdot \mathbf{B} $, where
\begin{align}
\mathbf{m} _{n\mathbf{k}} = - i \frac{e}{2 \hbar} \bra{\nabla_{\mathbf{k}} u _{n\mathbf{k}}} \times (\hat{H} - E _{n \mathbf{k}}  ) \ket{\nabla _{\mathbf{k}} u _{n\mathbf{k}}} . \label{OMM}
\end{align}
generically describes the rotation of a wave packet around its center of mass \cite{PhysRevB.53.7010, PhysRevB.59.14915}. Here, $E _{n \mathbf{k}} $ and $\ket{u _{n\mathbf{k}}}$ correspond to the unperturbed band structure. Both $\boldsymbol{\Omega} _{n\mathbf{k}}$ and $\mathbf{m} _{n\mathbf{k}}$ inherit the anisotropic character of the dispersion in the vicinity of the nodal line and must therefore be treated consistently within the gradient expansion.

In the absence of thermal and chemical-potential gradients, the electric current density is given by \cite{Ziman1972}
\begin{align}
\mathbf{J} = - e \sum _{n} \int \frac{d^{3}\mathbf{k}}{(2\pi)^{3}} \;  D ^{-1} _{n\mathbf{k}} \, \dot{\mathbf{r}} _{n} \; f _{n} , \label{Current}
\end{align}
where $f _{n}$ denotes the nonequilibrium distribution function. The expression in Eq.~(\ref{Current}) is written in terms of the full semiclassical velocity \eqref{rdot_decoupled} and incorporates the Berry-curvature correction to the phase-space measure through the factor $D _{n}$.

The distribution function is determined from the Boltzmann equation,
\begin{align}
\left( \frac{\partial}{\partial t} + \dot{\mathbf{r}}_{n}\cdot\nabla_{\mathbf{r}} + \dot{\mathbf{k}}_{n}\cdot\nabla_{\mathbf{k}} \right)f _{n} = - \frac{f _{n} - f ^{\mathrm{eq}} (\mathcal{E} _{n \mathbf{k}}) }{\tau}, \label{Boltzmann_Eq}
\end{align}
where we adopt the relaxation-time approximation with a constant scattering time $\tau$. The equilibrium distribution is given by $f ^{\mathrm{eq}} (\varepsilon) = [ 1 + e ^{\beta ( \varepsilon - \mu ) } ] ^{-1} $.

For homogeneous systems in steady state, Eq.~(\ref{Boltzmann_Eq}) can be solved iteratively. To leading order in the electric field, one finds
\begin{align}
f _{n} = f ^{\mathrm{eq}} (\mathcal{E} _{n \mathbf{k}}) + e \tau D _{n\mathbf{k}} \; \mathbf{E} \cdot \left[ \mathbf{v} _{n} + \frac{e}{\hbar}  ( \mathbf{v} _{n} \cdot \boldsymbol{\Omega} _{n} ) \; \mathbf{B} \right] \frac{\partial f ^{\mathrm{eq}} (\mathcal{E} _{n \mathbf{k}})}{\partial \mathcal{E} _{n \mathbf{k}} } . \label{Boltzmann_Eq_Sol}
\end{align}
While Eq.~(\ref{Boltzmann_Eq_Sol}) captures the linear-response regime, nonlinear transport arises from higher-order corrections obtained through successive iterations of the Boltzmann equation, together with the intrinsic field dependence of the semiclassical dynamics. In nodal-line semimetals, these contributions are controlled by the geometry of the Fermi surface surrounding the nodal loop, rather than by topological charge associated with isolated band crossings. Consequently, second-order responses encode detailed information about the anisotropic distribution of Berry curvature and orbital magnetic moment along the Fermi surface. The present treatment assumes that magnetic-field effects do not lead to Landau-level quantization, so that transport is governed by semiclassical dynamics. {Within this regime, the semiclassical Boltzmann formalism has proven remarkably successful in describing geometric transport responses in topological semimetals, including Hall, planar Hall, and thermoelectric effects~\cite{Ipsita_Gapped_NLS,Haidar2025}.}

Substituting the semiclassical velocity \eqref{rdot_decoupled} and the solution of the Boltzmann equation \eqref{Boltzmann_Eq_Sol} into the current expression \eqref{Current}, the electric current can be naturally decomposed into distinct contributions according to their physical origin. First, the equilibrium contribution in the absence of an electric field reads
\begin{align}
\mathbf{J} ^{\mathrm{eq}} = - e \sum _{n} \int \frac{d ^{3}\mathbf{k}}{( 2 \pi ) ^{3} } \; \boldsymbol{\mathcal{V}} _{n\mathbf{k}} \; f ^{\mathrm{eq}} ( \mathcal{E} _{n \mathbf{k}} ) ,
\end{align}
which vanishes identically in homogeneous systems due to the absence of a net current in equilibrium. We therefore discard this term in what follows. In the above expression we introduced the generalized velocity
\begin{align}
\boldsymbol{\mathcal{V}}_{n\mathbf{k}} = \mathbf{v} _{n\mathbf{k}} + \frac{e}{\hbar} \left( \mathbf{v} _{n\mathbf{k}} \cdot \boldsymbol{\Omega} _{n\mathbf{k}} \right) \mathbf{B} , \label{generalized_velocity}
\end{align}
which incorporates the Berry-curvature-induced correction to the transport velocity.

The remaining current can be separated into two physically distinct parts. The first corresponds to the intrinsic Hall response arising from the anomalous velocity term,
\begin{align}
\mathbf{J} ^{\mathrm{Hall}}  = - \frac{e ^{2}}{\hbar} \, \mathbf{E} \times \sum _{n} \int \frac{d ^{3}\mathbf{k}}{( 2 \pi ) ^{3}} \; \boldsymbol{\Omega} _{n \mathbf{k}} \; f ^{\mathrm{eq}}(\mathcal{E} _{n \mathbf{k}}) . \label{J_intrinsic}
\end{align}
This contribution is independent of the relaxation time $\tau$ and therefore represents a purely intrinsic effect determined by the Berry curvature of the occupied states. In nodal-line semimetals, where the Berry curvature is distributed along the nodal loop rather than concentrated at isolated points, this term reflects the geometry of the Fermi surface surrounding the line node. Its magnitude and symmetry are controlled by the anisotropic structure of $\boldsymbol{\Omega}_{n\mathbf{k}}$ and by the field dependence of the dispersion through the orbital magnetic moment. {Similar geometric mechanisms have recently been discussed in the context of Hall and magnetothermal transport in topological semimetals~\cite{Haidar2025}.}

The second contribution originates from the nonequilibrium correction to the distribution function and corresponds to the Fermi-surface transport current,
\begin{align}
\mathbf{J} ^{\mathrm{FS}}  = - e ^{2} \tau \,  \sum _{n} \int \frac{d ^{3}\mathbf{k}}{( 2 \pi ) ^{3}} \; D _{n\mathbf{k}} \, ( \mathbf{E} \cdot \boldsymbol{\mathcal{V}} _{n\mathbf{k}} ) \, \boldsymbol{\mathcal{V}} _{n\mathbf{k}} \, \frac{\partial f^{\mathrm{eq}}(\mathcal{E} _{n \mathbf{k}})}{\partial \mathcal{E} _{n \mathbf{k}} } . \label{J_FS}
\end{align}
This term depends explicitly on the scattering time and therefore describes dissipative transport. It contains all Fermi-surface contributions, including both conventional Drude conductivity and magnetic-field-induced corrections arising from the Berry curvature and the orbital magnetic moment.

The phase-space factor $D _{n\mathbf{k}}$ and the generalized velocity $\boldsymbol{\mathcal{V}}_{n\mathbf{k}}$ incorporate geometric effects associated with $\boldsymbol{\Omega}_{n\mathbf{k}}$, while the field dependence of the band energy $\mathcal{E} _{n \mathbf{k}} = E _{n\mathbf{k}}  - \mathbf{m} _{n\mathbf{k}} \cdot \mathbf{B} $ introduces additional corrections through both the velocity and the derivative of the distribution function. As a result, Eq.~(\ref{J_FS}) contains contributions of different orders in the magnetic field, which can be systematically classified into orbital-magnetic-moment, Berry-curvature, and mixed sectors. {This decomposition provides a convenient framework for comparing the present nonperturbative formulation with previous perturbative descriptions of Hall and planar Hall transport in gapped nodal-line semimetals~\cite{Ipsita_Gapped_NLS,PhysRevB.108.085120}.}

\section{Nodal-line semimetal model}
\label{Model_NLSM}

We now introduce a minimal two-band model describing a nodal-line semimetal characterized by a closed band-touching loop in momentum space. In contrast to Weyl semimetals, where band crossings occur at isolated points acting as monopoles of Berry curvature \cite{RevModPhys.90.015001, 10.1146/annurev031016-025458}, nodal-line systems exhibit an extended degeneracy forming a one-dimensional manifold, typically protected by crystalline symmetries or approximate chiral symmetry \cite{Yang31122022}.

The low-energy Hamiltonian is taken as \cite{PhysRevLett.115.026403, PhysRevB.92.081201}
\begin{align}
\hat{H} (\mathbf{k}) = \hbar v q_{\rho} \sigma _{x} + \hbar v k_{z} \sigma_{y} + m   \sigma _{z}  , \label{Hamiltonian_NLSM}
\end{align}
where $q _{\rho} = k _{\rho} - k _{0}$, with $k _{\rho} = \sqrt{k _{x} ^{2} + k _{y} ^{2} }$ the radial coordinate in momentum space. The parameter $k _{0}$ sets the radius of the nodal ring in the $k _{x}$-$k _{y}$ plane, while $m$ controls the gap opening away from the nodal line \cite{10.1063/5.0030200}. In the limit $m \to 0$, the conduction and valence bands touch along the circle defined by $k _{z} = 0$ and $k _{\rho} = k _{0} $.

The energy spectrum follows directly from Eq.~(\ref{Hamiltonian_NLSM}) as
\begin{align}
E _{n\mathbf{k}}  = n  \lambda _{\mathbf{k}} , \qquad \lambda _{\mathbf{k}} = \sqrt{ m ^{2}  + ( \hbar v ) ^{2} \bigl( k _{z} ^{2} + q _{\rho} ^{2} \bigr) } , \qquad n = \pm 1 , \label{Dispersion_NLSM}
\end{align}
showing that the dispersion is isotropic in the $(k _{z},q _{\rho})$ plane, i.e., in the directions perpendicular to the nodal ring. The resulting energy landscape consists of a toroidal structure in momentum space, with low-energy excitations localized around the nodal loop. The corresponding band velocity is obtained as the gradient of the dispersion,
\begin{align}
\boldsymbol{\upsilon} _{n \mathbf{k}} = \frac{1}{\hbar} \nabla _{\mathbf{k}} E _{n\mathbf{k}} = \frac{n v ^{2} \hbar}{\lambda _{\mathbf{k}} }  \left( q _{\rho} \, \hat{\mathbf{e}} _{\rho} + k _{z} \, \hat{\mathbf{e}} _{z} \right) , \label{Band_Velocity_NLSM}
\end{align}
where $\hat{\mathbf{e}} _{\rho} = \cos \varphi \, \hat{\mathbf{e}} _{x} +  \sin \varphi \, \hat{\mathbf{e}} _{y}$ is the radial unit vector in the $k _{x}$-$k _{y}$ plane.

Despite the absence of isolated monopoles in momentum space, the Berry curvature of this model is nonvanishing and exhibits a characteristic circulating structure around the nodal ring. From the Bloch eigenstates of Eq.~(\ref{Hamiltonian_NLSM}), one obtains
\begin{align}
\boldsymbol{\Omega} _{n \mathbf{k}}  = - \frac{n m v ^{2} \hbar ^{2}}{2 \lambda ^{3} _{\mathbf{k}} } \, \hat{\mathbf{e}} _{\varphi} , \label{Berry_Curvature_NLSM}
\end{align}
where the azimuthal unit vector is defined as $\hat{\mathbf{e}} _{\varphi} = - \sin \varphi \, \hat{\mathbf{e}} _{x} + \cos \varphi \, \hat{\mathbf{e}} _{y}$. This structure reflects the fact that the Berry curvature forms closed loops encircling the nodal line, rather than emanating radially from a point as in Weyl systems. The magnitude of $\boldsymbol{\Omega} _{n \mathbf{k}}$ is controlled by the mass parameter $m$, vanishing in the limit $m \to 0$, where the system recovers an exact band degeneracy along the nodal ring.

The orbital magnetic moment (\ref{OMM}) is found to be
\begin{align}
\mathbf{m} _{n \mathbf{k}}  = n  \lambda _{\mathbf{k}} \, \frac{ e }{\hbar} \, \boldsymbol{\Omega} _{n \mathbf{k}} .
\end{align}
As in generic two-band models, the orbital magnetic moment is proportional to the Berry curvature weighted by the band energy scale. Several important features follow directly from the above expressions. First, the Berry curvature and orbital magnetic moment are purely azimuthal, reflecting the underlying cylindrical symmetry of the nodal-line geometry \cite{PhysRevB.98.155125}. Second, both quantities are strongly peaked near the nodal ring, where $\lambda _{\mathbf{k}}$ is minimal, and decay rapidly away from it. Third, in contrast to Weyl semimetals, where the Berry curvature acts as a source or sink in momentum space, here it circulates around the nodal loop, leading to qualitatively different transport signatures.

In particular, the absence of a net monopole charge implies that linear anomalous Hall responses vanish in centrosymmetric configurations \cite{PhysRevLett.115.216806}, while nonlinear and magnetic-field-induced responses can remain finite due to the nontrivial distribution of Berry curvature and orbital magnetic moment around the nodal line. These features will play a central role in the analysis of magnetotransport phenomena developed in the following sections.

\section{Electric current and transport contributions}
\label{Current_Section}

In this section, we evaluate the electric current generated by the semiclassical dynamics introduced above and identify the distinct physical mechanisms contributing to transport in nodal-line semimetals. Two qualitatively different mechanisms can be distinguished. The first is an intrinsic contribution arising from the anomalous velocity induced by the Berry curvature, which is present already in equilibrium and does not depend on scattering processes. The second originates from the nonequilibrium correction to the distribution function and is associated with quasiparticle dynamics at the Fermi surface \cite{PhysRevLett.93.206602, PhysRevLett.97.026603}. This latter contribution depends explicitly on the relaxation time and encodes the dissipative transport response.

In nodal-line semimetals, this separation acquires particular significance. While the intrinsic response is governed by the geometric distribution of Berry curvature along the nodal loop, the Fermi-surface contribution reflects the anisotropic structure of the electronic states in its vicinity. Moreover, both contributions are modified by the presence of the orbital magnetic moment, which introduces an explicit magnetic-field dependence in the band dispersion and consequently affects both the velocity and the phase-space measure. In the following, we analyze these two contributions separately.

\subsection{Intrinsic Hall response from anomalous velocity}
\label{subsec:intrinsic_hall}

We first consider the intrinsic contribution to the current arising from the anomalous velocity term in the semiclassical equations of motion. This contribution is independent of the relaxation time and is entirely determined by the Berry curvature of the occupied states. Using the explicit form of the Berry curvature (\ref{Berry_Curvature_NLSM}) the intrinsic Hall current (\ref{J_intrinsic}) can be written as
\begin{align}
    \mathbf{J} ^{\mathrm{Hall}} &= \frac{m e ^{2} v ^{2} \hbar}{2}  \, \mathbf{E} \times \sum _{n}  \int \frac{d^{3}\mathbf{k}}{(2\pi)^{3}} \; n \frac{ \hat{\mathbf{e}} _{\phi} }{E _{n \mathbf{k}} ^{3} }   \; \Theta \left( \mu - E _{n \mathbf{k}} - \frac{e mv ^{2} \hbar  }{2} \,    \frac{ \mathbf{B} \cdot  \hat{\mathbf{e}} _{\phi} }{E _{n \mathbf{k}} ^{2} }  \right) , \label{J_intrinsic}
\end{align}
where $\Theta (x)$ is the Heaviside step function. For nodal-line semimetals, the Berry curvature is purely azimuthal and circulates around the nodal loop. As a consequence, in the absence of a magnetic field the argument of the step function is independent of the azimuthal angle, and the angular integral reduces to $\int_{0}^{2\pi} d\phi\, \hat{\mathbf e}_{\phi}=0$, so that the intrinsic Hall contribution vanishes identically by symmetry \cite{PhysRevB.97.161113, PhysRevB.98.155125}. The situation changes once the orbital magnetic moment is included, since it introduces an explicit angular dependence through $\mathbf{B}\cdot \hat{\mathbf e}_{\phi}$ in the occupation function. In this way, the magnetic field acts as an angular selector, breaking the exact cancellation of the azimuthal integral and allowing for a finite response. Although both bands may in principle contribute, for $\mu>m$ the valence band ($n=-1$) remains fully occupied throughout the integration domain, and in the weak-field regime the argument of the step function stays positive for all $\phi$, so that its contribution still vanishes by symmetry. By contrast, the conduction band ($n=+1$) is only partially occupied, and the field-induced angular selection leads to a finite intrinsic Hall current. Therefore, the response is entirely determined by the conduction band, and in the following we set $n=+1$.

At this stage, it is convenient to introduce a toroidal coordinate system adapted to the nodal-line geometry of the semimetal. In this parametrization, the momentum components are written as \cite{PhysRevLett.119.147402}
\begin{align}
    k _{x} &= (k _{0} + r \cos \theta ) \cos \phi ,  \notag \\ k _{y} &= (k _{0} + r \cos \theta ) \sin \phi ,  \notag \\ k _{z} &=  r \sin \theta , \label{toroidal_coordinates}
\end{align}
where $k _{0}$ denotes the radius of the nodal ring in the $(k_x,k_y)$ plane. The radial coordinate $r \in [0, k _{0} ] $ measures the distance from the nodal line, $\theta \in [0, 2 \pi )$ parametrizes the transverse direction normal to the ring, and $\phi \in [0, 2 \pi )$ is the azimuthal angle along the nodal loop. With this choice, the energy dispersion becomes $E _{n \mathbf{k}} = n \lambda (r)$, where $\lambda (r) = \sqrt{ m ^{2} + (\hbar v r) ^{2} }$, while the momentum-space volume element acquires the Jacobian $J(r,\theta ) = r \, (k _{0} + r \cos \theta ) $. This coordinate system thus provides a natural framework to evaluate the Hall conductance. In this coordinate system, after integrating over $\theta$, equation (\ref{J_intrinsic}) for the conduction band becomes
\begin{align}
    \mathbf{J} ^{\mathrm{Hall}} &= \frac{e ^{2} }{\hbar}  \frac{m k _{0} }{16 \pi ^{2} }   \; \mathbf{E} \times \int _{m} ^{  \sqrt{ m ^{2} + (\hbar v k _{0} ) ^{2} }  } \frac{d \lambda}{\lambda ^{2}} \;  \int _{0} ^{2 \pi } d \phi  \; \hat{\mathbf{e}} _{\phi} \;  \Theta \left( \mu -   \lambda  -  \frac{e mv ^{2} \hbar  }{2} \,    \frac{ \mathbf{B} \cdot  \hat{\mathbf{e}} _{\phi} }{ \lambda ^{2}  }  \right) .  \label{J_intrinsic2}
\end{align}
The angular integral can be evaluated analytically (see Appendix~\ref{Angular_Integral}), yielding
\begin{align}
    \mathbf{J} ^{\mathrm{Hall}} &=   \frac{e ^{2} k _{0} }{4 \pi h}   \; \hat{\mathbf{B}} \times \mathbf{E}  \; \int _{m} ^{  \sqrt{ m ^{2} + (\hbar v k _{0} ) ^{2} }  } \frac{m  \, d \lambda}{\lambda ^{2}} \;  \sqrt{ 1 - p ^{2} ( \lambda ) } \;
    \Theta \bigl( 1 - | p ( \lambda ) | \bigr) , \qquad  p ( \lambda) = \frac{ 2 \lambda ^{2} ( \mu - \lambda ) }{ e mv ^{2} \hbar B } .   \label{J_intrinsic3}
\end{align}
To evaluate the corresponding integral, we introduce the change of variables $\xi = \lambda / m $ and use the dimensionless variables $\tilde{\mu} = \mu / m$, $\Delta = \hbar v k _{0} / m$,  $ b = B / B _{\mathrm{eff}}$, and $B _{\mathrm{eff}} = \frac{2 m ^{2}}{e  v ^{2} \hbar}$ is an effective magnetic field scale. Therefore, the intrinsic Hall response becomes
\begin{align}
    \mathbf{J} ^{\mathrm{Hall}} &=  \sigma ^{\mathrm{Hall}}  \; \hat{\mathbf{B}} \times \mathbf{E}  \; \Sigma (b,\Delta, \tilde{\mu}) .   \label{J_intrinsic4}
\end{align}
where $\sigma ^{\mathrm{Hall}}  = \frac{e ^{2} k _{0} }{4 \pi h} $ and
\begin{align}
    \Sigma (b,\Delta, \tilde{\mu}) = \frac{1}{b} \int _{1} ^{   \sqrt{ 1 + \Delta ^{2} }  } \frac{ d \xi }{\xi ^{2}} \;  \sqrt{ b ^{2} - \xi ^{4} ( \tilde{\mu} - \xi ) ^{2}  } \;    \Theta \bigl( b ^{2} - \xi ^{2}  | \tilde{\mu} - \xi | \bigr)   \label{f_function}
\end{align}
is a dimensionless function. Several important features follow from Eq.~(\ref{J_intrinsic4}). First, the current is always transverse to both the electric and magnetic fields, reflecting its Hall-like character. Second, the direction of the response is determined by $\hat{\mathbf{B}}$, indicating that the magnetic field selects a preferred axis that breaks the azimuthal symmetry of the nodal loop. Third, the magnitude of the response is controlled by the function $\Sigma$, which originates from the interplay between Berry curvature and orbital magnetic moment.

In order to obtain analytic insight into Eq.~(\ref{f_function}), we assume that the chemical potential lies in the conduction band, slightly above the gap, $\tilde{\mu} \gtrsim 1$, so that transport is dominated by extended states at the Fermi surface rather than by thermally activated carriers across the gap. We then consider the weak-field regime $b \ll 1$, corresponding to magnetic fields much smaller than the characteristic scale $B_{\mathrm{eff}}$. In this limit, the magnetic energy scale is much smaller than the Fermi energy, i.e. $B \ll \mu ^{2} / (e \hbar v ^{2} )$, so that Landau quantization can be neglected and the semiclassical description remains valid. Under these conditions, the integral in Eq.~(\ref{f_function}) is dominated by contributions close to the lower bound $\xi \sim 1$, which allows for a controlled expansion of the integrand in powers of $b$. The final result is $\Sigma (b,\Delta,\tilde\mu) \simeq \pi b / 2$, such that
\begin{align}
    \mathbf{J} ^{\mathrm{Hall}} =   \frac{e ^{2} k _{0} }{4 \pi h}   \; \boldsymbol{b} \times \mathbf{E}
\end{align}
where $\boldsymbol{b} = \mathbf{B} / B _{\mathrm{eff}}$. This result highlights a key qualitative difference with Weyl semimetals. While in Weyl systems the intrinsic Hall response is present already at zero magnetic field due to the monopolar structure of the Berry curvature, in nodal-line semimetals the intrinsic Hall current is induced only in the presence of a magnetic field. It therefore represents a field-activated geometric response, originating from the redistribution of Berry curvature on the Fermi surface due to orbital magnetic moment effects.

\begin{figure}
    \centering
    \includegraphics[width=0.47\linewidth]{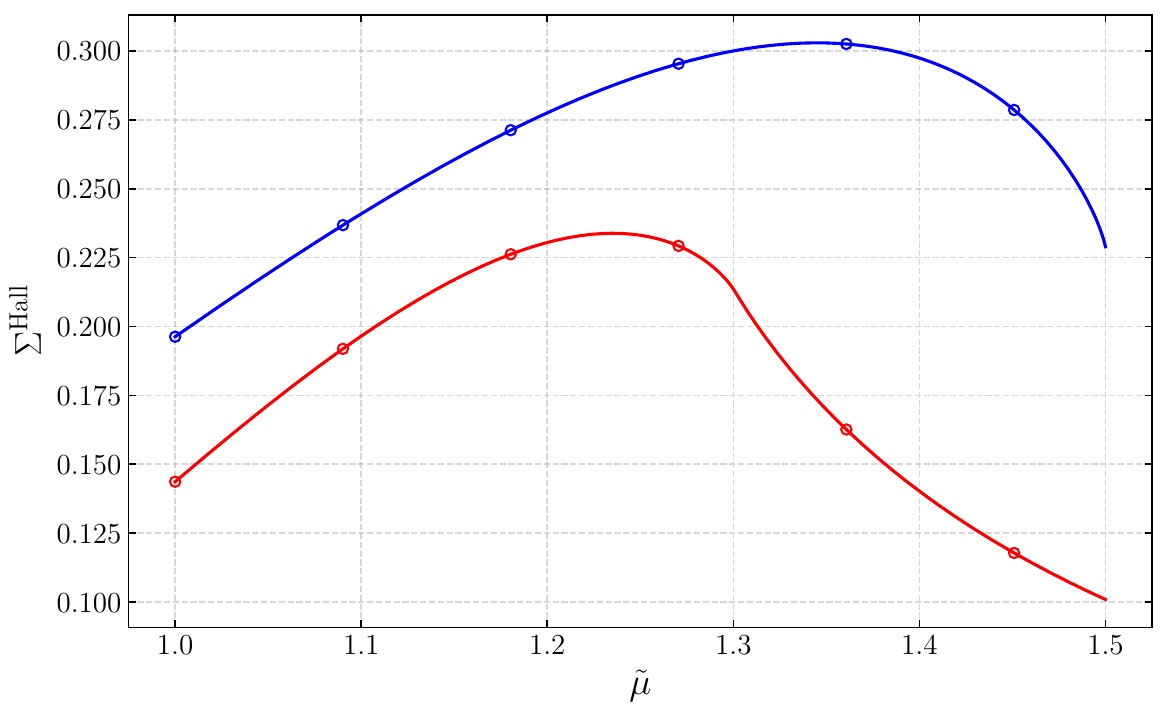}
    \includegraphics[width=0.47\linewidth]{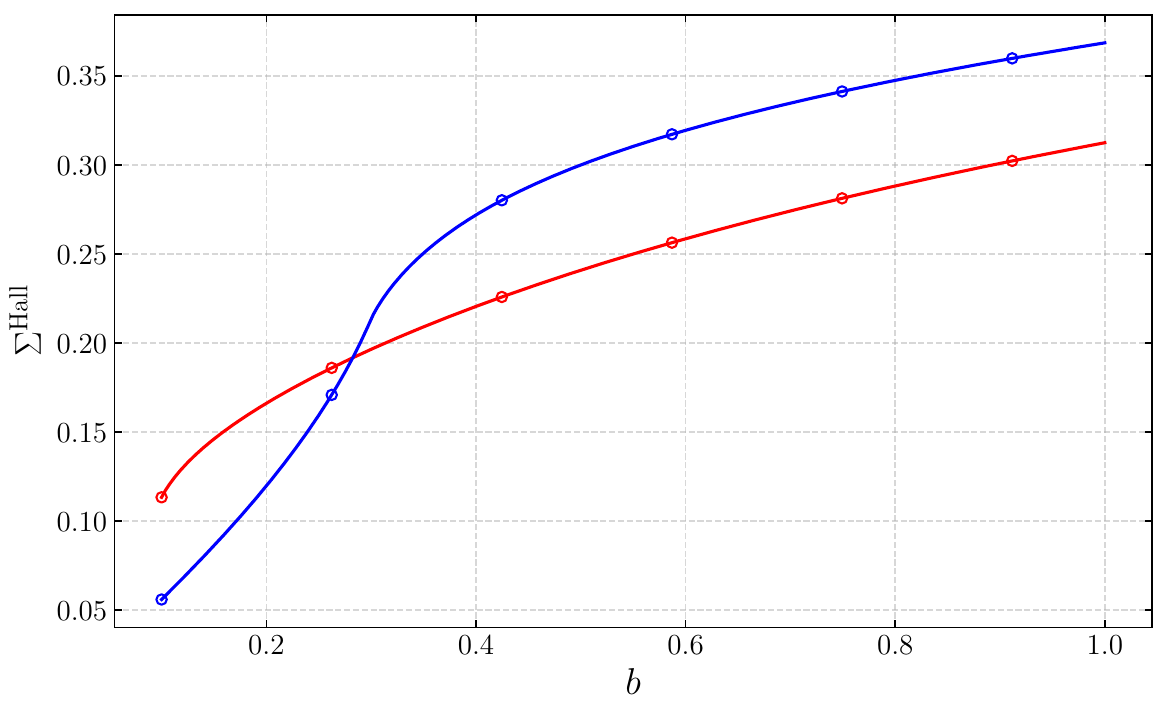}
    \caption{Dimensionless intrinsic Hall response $\Sigma(b,\Delta,\tilde{\mu})$ defined in Eq.~(\ref{f_function}).  Left panel: dependence on the rescaled chemical potential $\tilde{\mu}=\mu/m$ for two values of the magnetic field, $b=0.3$ (red) and $b=0.5$ (blue). The response increases close to the gap edge, reaches a maximum, and then decreases as the Fermi level moves deeper into the conduction band.  Right panel: magnetic-field dependence as a function of $b=B/B_{\mathrm{eff}}$ for fixed values $\tilde{\mu}=1.1$ (red) and $\tilde{\mu}=1.3$ (blue). The linear weak-field regime is followed by a pronounced nonlinear behavior associated with orbital-magnetic-moment effects and the field-induced restriction of the phase space.}  \label{Hall_plots}
\end{figure}

For the numerical calculations, we use parameters that mimic a SrAs$_{3}$-type nodal-line semimetal. This material crystallizes in a monoclinic lattice (space group $C2/c$), where the low-energy bands are mainly formed by hybridized As-$p$ and Sr-$d$ orbitals arranged in extended As$_6$ octahedral units. Experimental probes such as ARPES, together with first-principles studies, indicate that in the absence of spin-orbit coupling the system supports a single nodal ring located close to the Fermi energy, with only a small gap of a few meV induced once spin-orbit effects are included~\cite{Lv2018_SrAs3}.

Guided by these observations, we choose the parameter set $m = 5~\mathrm{meV}$, $v \simeq 3.2\times10^{5}~\mathrm{m/s}$, and $k_{0} \simeq 0.066~\text{\AA}^{-1}$ \cite{Lv2018_SrAs3}. These values determine the relevant dimensionless scales entering the problem. In particular,
\begin{align}
\Delta = \frac{\hbar v k_{0}}{m}
\end{align}
yields $\Delta \simeq 28$, showing that the nodal energy $\hbar v k_{0}$ largely excedes the gap scale. As a consequence, the low-energy dinamics is still governed by the nodal-line structure, with only a weak symmetry-breaking perturbation. The effect of the magnetic field is conveniently parametrized in terms of
\begin{align}
b = \frac{B}{B_{\mathrm{eff}}}, 
\qquad 
B_{\mathrm{eff}} = \frac{2 m^{2}}{e v^{2} \hbar},
\end{align}
which defines the natural field scale associated with orbital magnetic moment corrections. For the parameters above, one finds $B_{\mathrm{eff}} \simeq 0.74~\mathrm{T}$. Throughout this work we focus on the regime $b<1$, where the semiclassical description remains valid and Landau quantization can be safely neglected.

The numerical evaluation of the dimensionless function
$\Sigma(b,\Delta,\tilde{\mu})$ is presented in
Fig.~\ref{Hall_plots}, where we separately analyze its dependence on the
chemical potential and on the magnetic field. We consider two complementary
situations: (i) the dependence on the rescaled chemical potential
$\tilde{\mu}=\mu/m$ for fixed magnetic fields $b=0.3$ and $b=0.5$, and
(ii) the magnetic-field dependence for fixed chemical potentials
$\tilde{\mu}=1.1$ and $\tilde{\mu}=1.3$, both lying slightly above the gap.

We first consider the dependence on the chemical potential, shown in the
left panel of Fig.~\ref{Hall_plots}. Starting from the gap edge
($\tilde{\mu}=1$), the Hall response increases as additional conduction-band
states become available. Since the Berry curvature and the orbital magnetic
moment are strongly concentrated around the nodal ring, the newly occupied
states contribute efficiently to the intrinsic Hall current, producing a
rapid enhancement of $\Sigma$. This initial increase can also be understood
directly from Eq.~(\ref{f_function}). The Heaviside function restricts the
integration to the field-dependent window
$|\tilde{\mu}-\xi|\lesssim b/\xi^{2}$, which is partially truncated by the
lower integration limit $\xi=1$ when the chemical potential lies close to the
gap edge. As $\tilde{\mu}$ increases, this restriction becomes less severe,
allowing a larger portion of the geometrically active states around the nodal
ring to contribute to the Hall response.

The response reaches a well-defined maximum, whose magnitude increases with
the magnetic field. Beyond this point, however, the response decreases as the
chemical potential moves deeper into the conduction band. Although the
available phase space continues to increase, the center of the allowed
integration window is simultaneously displaced toward larger values of
$\xi$, where the geometric weight is progressively suppressed by the factor
$\xi^{-2}$. In addition, the field-dependent constraint imposed by the
Heaviside function continues to limit the active integration region.
Consequently, electronic states located farther from the nodal ring contribute
much less efficiently to the Hall current, causing $\Sigma$ to decrease. The
nonmonotonic behavior therefore reflects the competition between two opposing
mechanisms: the progressive opening of the available phase space as the Fermi
level moves away from the gap, and the simultaneous reduction of the
geometric weight associated with states farther from the nodal line.

The right panel displays the magnetic-field dependence. For sufficiently
small fields, the response increases linearly with $b$, in agreement with
the analytical weak-field result $\Sigma\simeq\pi b/2$. As the magnetic
field increases, however, pronounced deviations from linearity develop,
revealing the intrinsically nonperturbative character of the Hall response.
These deviations originate from the orbital magnetic moment, which modifies
the field-corrected energy and continuously reshapes the region of momentum
space satisfying the constraint imposed by the Heaviside function. An interesting feature
is the strong dependence on the position of the Fermi level. While the
response for $\tilde{\mu}=1.3$ increases more rapidly at very small fields,
the two curves cross at intermediate fields, after which the Hall response
becomes larger for $\tilde{\mu}=1.1$. This inversion demonstrates that the
field-induced redistribution of the geometrically active states depends
sensitively on the location of the Fermi surface relative to the nodal ring.
When the chemical potential lies closer to the gap, the orbital magnetic
moment is more effective in selecting the regions of the Fermi surface that
carry the largest Berry curvature, thereby producing a stronger enhancement
of the Hall current at moderate magnetic fields.

Overall, Fig.~\ref{Hall_plots} demonstrates that the intrinsic Hall response
in gapped nodal-line semimetals is simultaneously controlled by the magnetic
field and by the position of the Fermi level. The response is activated by
the magnetic field, reaches its largest values when the chemical potential
remains close to the nodal ring, and exhibits clear nonlinear signatures that
cannot be captured within a perturbative expansion in powers of $B$.

\subsection{Fermi-surface transport current}
\label{subsec:FS_current}

We now turn to the contribution associated with the nonequilibrium correction to the distribution function. This term describes transport processes occurring at the Fermi surface and depends explicitly on the scattering time. It includes both the conventional Drude response and magnetic-field-induced corrections arising from Berry curvature and orbital magnetic moment effects. Using the result of Section \ref{Model_NLSM} the Fermi-surface transport current can be written as
\begin{align}
\mathbf{J} ^{\mathrm{FS}}  = e ^{2} \tau \,  \sum _{n} \int \frac{d ^{3}\mathbf{k}}{( 2 \pi ) ^{3}} \; D _{n\mathbf{k}} \, ( \mathbf{E} \cdot \boldsymbol{\mathcal{V}} _{n\mathbf{k}} ) \, \boldsymbol{\mathcal{V}} _{n\mathbf{k}} \; \delta \left( \mu - E _{n \mathbf{k}} - \frac{e mv ^{2} \hbar  }{2} \,    \frac{ \mathbf{B} \cdot  \hat{\mathbf{e}} _{\phi} }{E _{n \mathbf{k}} ^{2} }  \right) , \label{J_FS2}
\end{align}
with generalized velocity
\begin{align}
\boldsymbol{\mathcal V} _{n\mathbf{k}} &= \left( \frac{ \hbar v ^{2}}{ E _{n \mathbf{k}} } - e m \hbar ^{2} v ^{4}\frac{ \mathbf{B} \cdot \hat{\mathbf{e}} _{\phi} }{ E _{n \mathbf{k}} ^{4}} \right) \left( q _{\rho}\, \hat{\mathbf{k}} _{\parallel} + k _{z} \hat{\mathbf{e}} _{z} \right) - \frac{e m v ^{2}}{2} \frac{ \mathbf{B} \cdot \hat{\mathbf{k}} _{\parallel} }{k _{\rho} E _{n \mathbf{k}} ^{2}} \; \hat{\mathbf{e}} _{\phi} + \frac{  e^{2} m^{2} v^{4}\hbar}{4 k _{\rho} E _{n \mathbf{k}} ^{5}} \, (\mathbf{B} \cdot \hat{\mathbf{k}} _{\parallel} ) \, \mathbf{B} , \label{Generalized_velocity_NLSM}
\end{align}
where $\hat{\mathbf{k}} _{\parallel} = \cos \phi \, \hat{\mathbf{e}} _{x} + \sin \phi \, \hat{\mathbf{e}} _{y}$ and $\hat{\mathbf{e}} _{\phi} = \hat{\mathbf{e}} _{z} \times \hat{\mathbf{k}} _{\parallel}$. The Fermi-surface conductivity tensor associated with Eq.~(\ref{J_FS2}) can be written as
\begin{align}
\sigma_{ij}^{\mathrm{FS}} = e ^{2} \tau \int \frac{d ^{3} \mathbf{k}}{( 2 \pi ) ^{3} } \; D _{\mathbf{k}} \, \mathcal{V} _{i} \mathcal{V} _{j} \, \delta \left( \mu - E _{\mathbf{k}} - \frac{e m v ^{2} \hbar }{ 2 E _{\mathbf{k}} ^{2} } \,\mathbf{B} \cdot \hat{\mathbf{e}} _{\phi} \right).
\label{sigma_FS_main_start}
\end{align}
The angular integrations can be carried out analytically, as detailed in Appendix~\ref{App_FS_current}, leading to a reduced one-dimensional representation in terms of the radial variable $\lambda$. The resulting conductivity tensor takes the form
\begin{align}
\sigma_{ij}^{\mathrm{FS}}
=
\sigma_{\parallel}^{\mathrm{FS}} \,\hat B_i \hat B_j
+
\sigma_{\perp}^{\mathrm{FS}}\, (\hat B_{\perp})_i (\hat B_{\perp})_j
+
\sigma_z^{\mathrm{FS}}\,\delta_{iz}\delta_{jz},
\label{sigma_tensor_decomposition}
\end{align}
where $\hat{\mathbf B}_{\perp}=\hat{\mathbf e}_z \times \hat{\mathbf B}$. The scalar coefficients are given by the exact radial integrals
\begin{align}
\sigma_{\parallel}^{\mathrm{FS}}
&=
\sigma ^{\mathrm{Hall}} \; \frac{2 \tau m }{\hbar}
\int_{1}^{\sqrt{1+\Delta^2}} 
\frac{ \sqrt{1-\psi^2 } \, d\lambda }{b (2\lambda-\tilde\mu)}
\Bigg[
(\lambda^2-1) (3\lambda-2\tilde\mu)^2
+
\frac{2b^2/\Delta}{\sqrt{1+\Delta^2-\lambda^2}}
\left(\psi+\frac{b}{\lambda^3}\right)^2
\Bigg],
\label{sigma_parallel_exact}
\\[6pt]
\sigma_{\perp}^{\mathrm{FS}}
&=
\sigma ^{\mathrm{Hall}} \; \frac{2 \tau m }{\hbar}
\int_{1}^{\sqrt{1+\Delta^2}} 
\frac{d\lambda}{b (2\lambda-\tilde\mu)\sqrt{1-\psi^2 }}
\Bigg[
(\lambda^2-1)(3\lambda-2\tilde\mu)^2\psi^2
+
\frac{2b^2/\Delta}{\sqrt{1+\Delta^2-\lambda^2}}
(1-\psi^2)^2
\Bigg],
\label{sigma_perp_exact}
\\[6pt]
\sigma_{z}^{\mathrm{FS}}
&=
\sigma ^{\mathrm{Hall}} \; \frac{2 \tau m }{\hbar}
\int_{1}^{\sqrt{1+\Delta^2}} 
\frac{d\lambda}{b (2\lambda-\tilde\mu)\sqrt{1-\psi^2 }}
\left[ (\lambda^2-1) (3\lambda-2\tilde\mu)^2
\right],
\label{sigma_z_exact}
\end{align}
where $\psi(\lambda)=\lambda^2(\lambda-\tilde\mu)/b$. These expressions are fully nonperturbative in the magnetic field and incorporate the combined effects of Berry curvature, orbital magnetic moment, and the field-dependent phase-space factor. In particular, the magnetic field enters both through the integration kernel and through the kinematic constraint imposed by the Dirac delta function, leading to a highly nontrivial deformation of the Fermi surface.

In the weak-field regime $b\ll 1$, the constraint $|\psi(\lambda)|\le 1$ restricts the integral to a narrow region around $\lambda=\tilde\mu$, allowing for a controlled expansion. As shown in Appendix~\ref{App_FS_current}, this leads to the following analytical expressions:
\begin{align}
\sigma_{z}^{\mathrm{FS}}
&=
\sigma ^{\mathrm{Hall}} \; \frac{2 \pi \tau m }{\hbar}
\left[
\frac{\tilde\mu^2-1}{\tilde\mu}
+
\frac{b^2}{2}\frac{2-5\tilde\mu^2}{\tilde\mu^7}
\right]
+
\mathcal{O}(b^3),
\label{sigma_z_main_weak}
\\[6pt]
\sigma_{\parallel}^{\mathrm{FS}}
&=
\sigma ^{\mathrm{Hall}} \; \frac{\pi \tau m }{\hbar}
\left[
\frac{\tilde\mu^2-1}{\tilde\mu}
+
\frac{b^2}{4}\frac{2-5\tilde\mu^2}{\tilde\mu^7}
+
\frac{b^2}{2\,\Delta\,\tilde\mu^3\sqrt{1+\Delta^2-\tilde\mu^2}}
\right]
+
\mathcal{O}(b^3),
\label{sigma_parallel_main_weak}
\\[6pt]
\sigma_{\perp}^{\mathrm{FS}}
&=
\sigma ^{\mathrm{Hall}} \; \frac{\pi \tau m }{\hbar}
\left[
\frac{\tilde\mu^2-1}{\tilde\mu}
+
\frac{3b^2}{4}\frac{2-5\tilde\mu^2}{\tilde\mu^7}
+
\frac{3b^2}{2\,\Delta\,\tilde\mu^3\sqrt{1+\Delta^2-\tilde\mu^2}}
\right]
+
\mathcal{O}(b^3).
\label{sigma_perp_main_weak}
\end{align}
Several important features follow from these results. First, the leading contribution is isotropic,
\begin{align}
\sigma_{\parallel}^{\mathrm{FS}}
=
\sigma_{\perp}^{\mathrm{FS}}
=
\sigma ^{\mathrm{Hall}} \; \frac{\pi \tau m }{\hbar}
\frac{\tilde\mu^2-1}{\tilde\mu}
+
\mathcal{O}(b^2),
\label{sigma_FS_isotropic}
\end{align}
recovering a Drude-like behavior controlled by the Fermi surface. Second, anisotropy appears only at order $b^2$, reflecting the fact that magnetic-field-induced distortions of the Fermi surface enter at subleading order in this contribution. Finally, the presence of the additional term proportional to $(\Delta\,\sqrt{1+\Delta^2-\tilde\mu^2})^{-1}$ reveals a nontrivial dependence on the band geometry, which has no counterpart in conventional isotropic metals. These results provide a controlled analytical benchmark for the full nonperturbative expressions in Eqs.~(\ref{sigma_parallel_exact})-(\ref{sigma_z_exact}), and will be used below to interpret the numerical evaluation of the transport coefficients.

\begin{figure}
    \centering
    \includegraphics[width=0.47\linewidth]{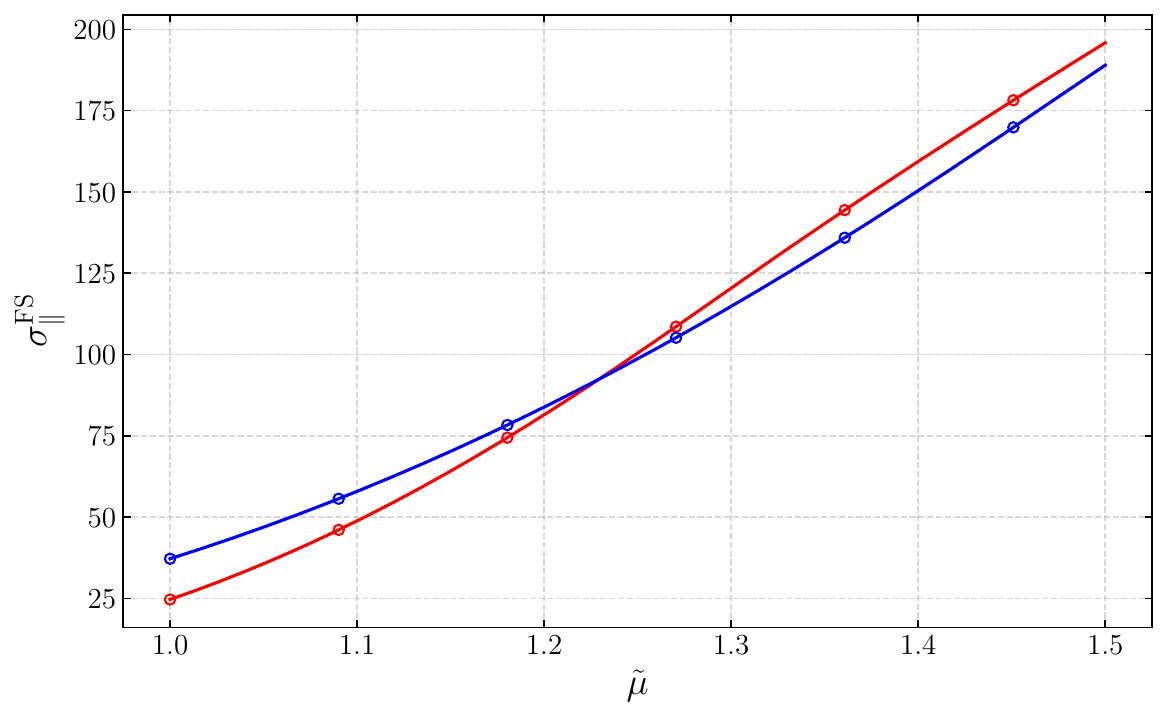}
    \includegraphics[width=0.47\linewidth]{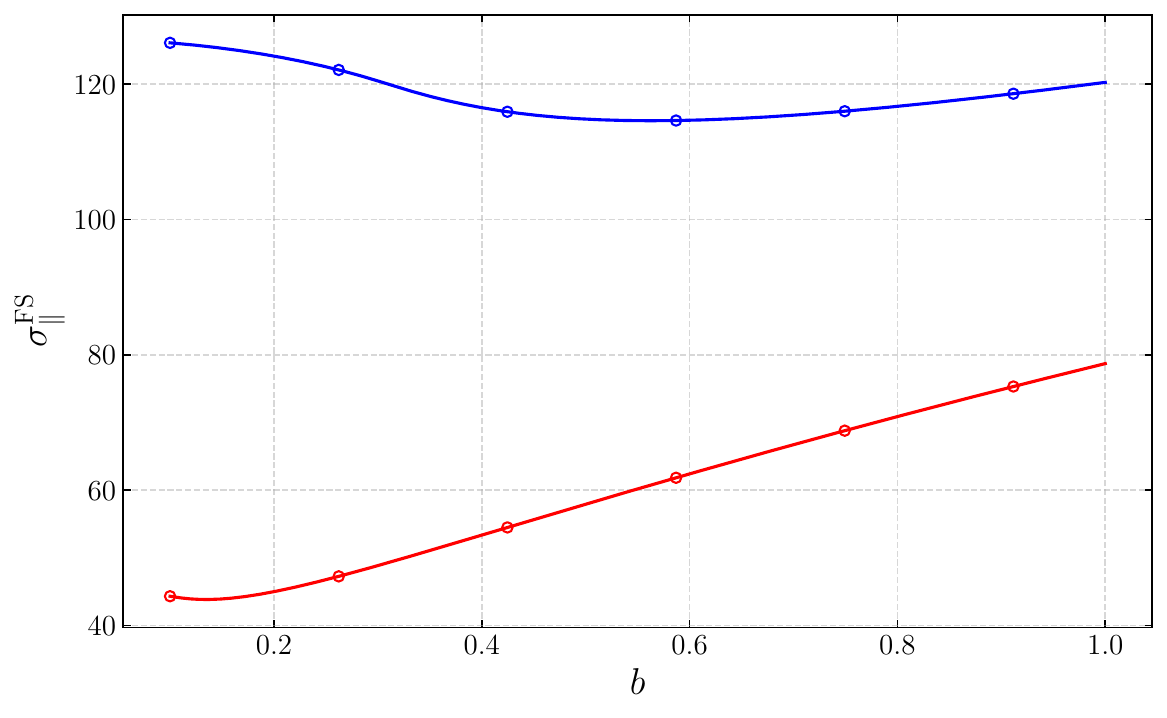}
    \caption{
Dimensionless longitudinal Fermi-surface conductivity
$\sigma_{\parallel}^{\mathrm{FS}}$, defined by Eq.~(\ref{sigma_parallel_exact}).
Left panel: dependence on the rescaled chemical potential
$\tilde{\mu}=\mu/m$ for two magnetic-field strengths,
$b=0.3$ (red) and $b=0.5$ (blue). The conductivity increases with carrier
density, while the magnetic field enhances the response mainly close to the
gap edge. Right panel: magnetic-field dependence for
$\tilde{\mu}=1.1$ (red) and $\tilde{\mu}=1.3$ (blue). The weak-field behavior
is followed by a pronounced nonlinear evolution resulting from the combined
effects of the orbital magnetic moment and the magnetic-field-induced
deformation of the Fermi surface.
}  \label{Sigma_parallel}
\end{figure}

\begin{figure}
    \centering
    \includegraphics[width=0.47\linewidth]{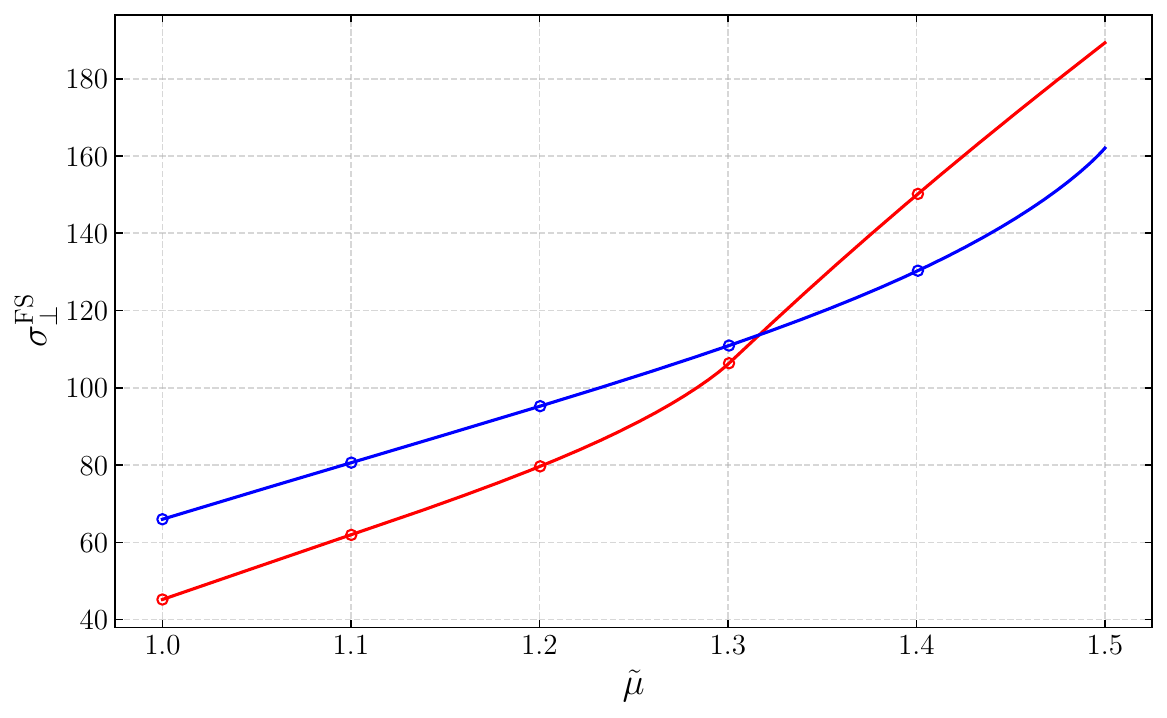}
    \includegraphics[width=0.47\linewidth]{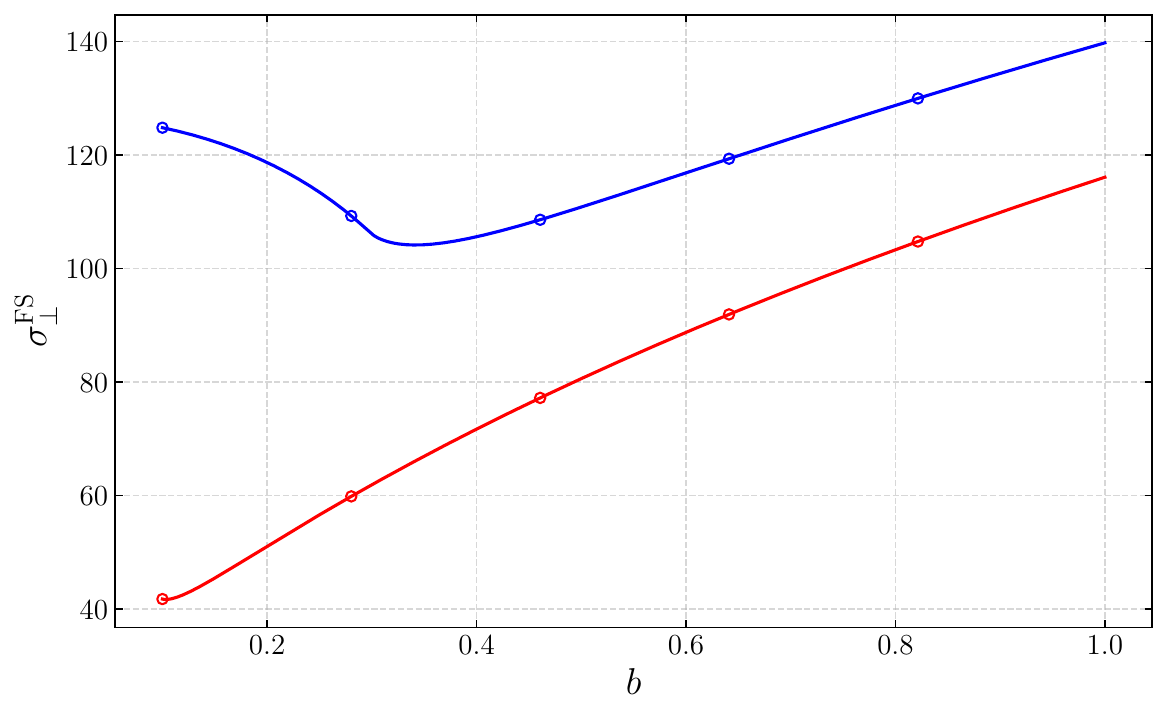}
    \caption{
Dimensionless transverse Fermi-surface conductivity
$\sigma_{\perp}^{\mathrm{FS}}$, defined by Eq.~(\ref{sigma_perp_exact}).
Left panel: dependence on the rescaled chemical potential
$\tilde{\mu}=\mu/m$ for $b=0.3$ (red) and $b=0.5$ (blue). The conductivity
grows monotonically with the chemical potential, while the magnetic field
induces a marked anisotropic enhancement close to the gap edge. Right panel:
magnetic-field dependence for $\tilde{\mu}=1.1$ (red) and
$\tilde{\mu}=1.3$ (blue). The nonmonotonic evolution reflects the progressive
competition between the field-induced deformation of the Fermi surface and the
redistribution of geometrically active states contributing to transport.
}  \label{Sigma_perp}
\end{figure}

\begin{figure}
    \centering
    \includegraphics[width=0.47\linewidth]{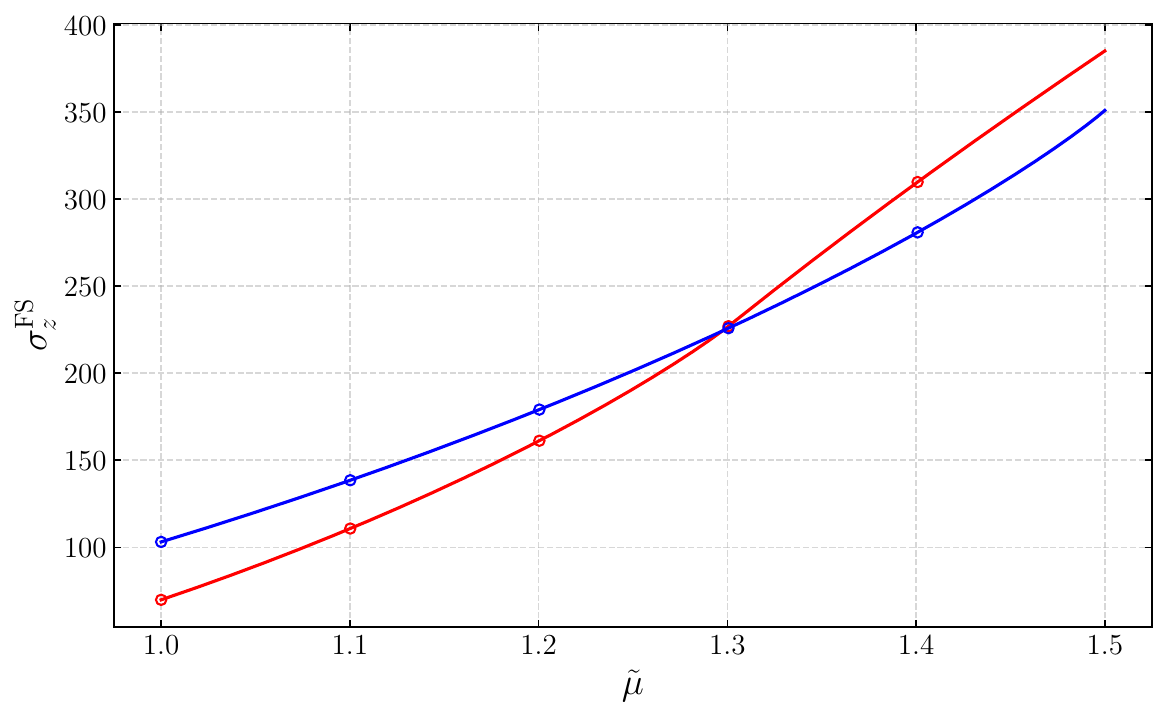}
    \includegraphics[width=0.47\linewidth]{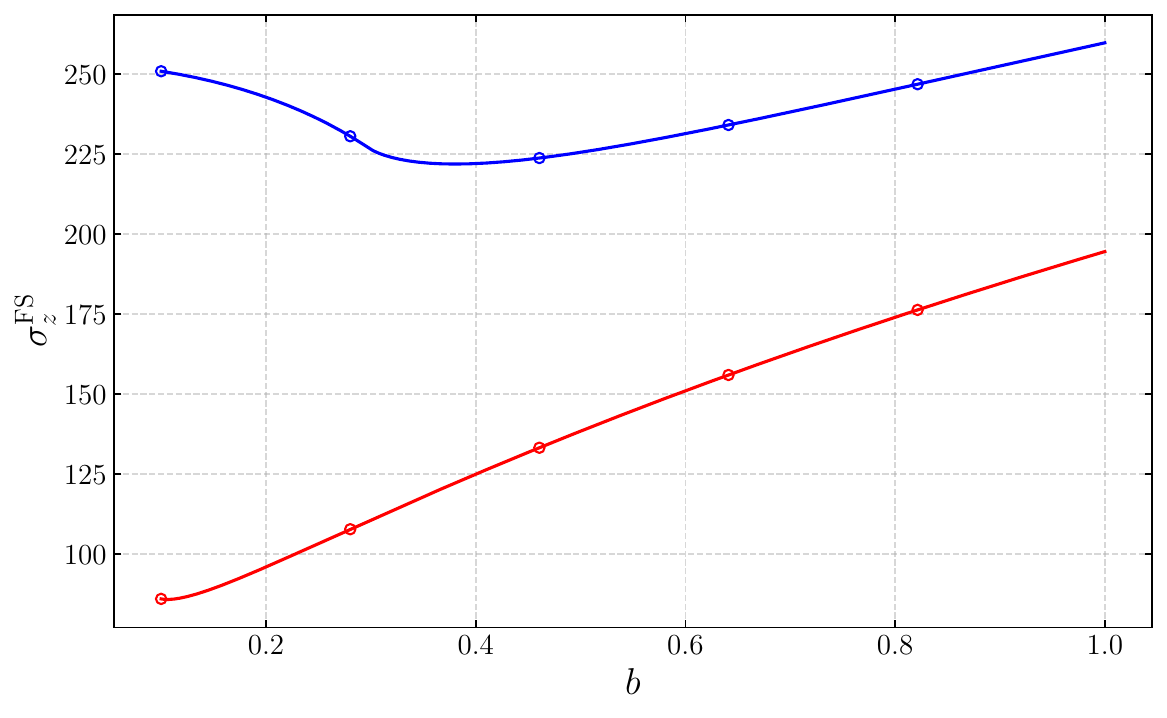}
    \caption{
Dimensionless out-of-plane Fermi-surface conductivity
$\sigma_{z}^{\mathrm{FS}}$, defined by Eq.~(\ref{sigma_z_exact}).
Left panel: dependence on the rescaled chemical potential
$\tilde{\mu}=\mu/m$ for magnetic fields
$b=0.3$ (red) and $b=0.5$ (blue). The conductivity increases steadily as the
Fermi level moves away from the gap, with magnetic-field effects being more
pronounced near the band edge. Right panel: dependence on the magnetic field
for $\tilde{\mu}=1.1$ (red) and $\tilde{\mu}=1.3$ (blue). Deviations from the
weak-field regime become evident at intermediate fields, illustrating the
nonperturbative influence of the orbital magnetic moment on Fermi-surface
transport.
}  \label{Sigma_z}
\end{figure}

Using the material parameters introduced in the previous subsection together with a representative relaxation time $\tau=10^{-14}\,\mathrm{s}$, we numerically evaluate the exact nonperturbative expressions in Eqs.~(\ref{sigma_parallel_exact})-(\ref{sigma_z_exact}). The resulting Fermi-surface conductivities are shown in Figs.~\ref{Sigma_parallel}-\ref{Sigma_z}, where the left panels display the dependence on the rescaled chemical potential at fixed magnetic field, while the right panels show the magnetic-field dependence for fixed values of the chemical potential.

The numerical evaluation of the Fermi-surface conductivities is shown in
Figs.~\ref{Sigma_parallel}-\ref{Sigma_z}. In all cases, the left panels
display the dependence on the rescaled chemical potential $\tilde{\mu}$ at
fixed magnetic field, while the right panels show the magnetic-field
dependence at fixed $\tilde{\mu}$. The three components exhibit a common
overall trend: the conductivity increases as the chemical potential moves
away from the gap edge. This behavior is already anticipated by the leading
weak-field contribution in Eq.~(\ref{sigma_FS_isotropic}), which is
proportional to $(\tilde{\mu}^{2}-1)/\tilde{\mu}$ and therefore grows with
the size of the Fermi surface.

The left panels of Figs.~\ref{Sigma_parallel}-\ref{Sigma_z} show that this
Drude-like increase with $\tilde{\mu}$ remains the dominant tendency beyond
the weak-field regime. However, the two magnetic-field curves are not simply
related by an overall rescaling. Close to the gap edge, the larger field
$b=0.5$ enhances the conductivity with respect to $b=0.3$, indicating that
orbital-magnetic-moment effects are more efficient when the Fermi surface is
located near the nodal ring, where the geometric quantities are largest. At
larger values of $\tilde{\mu}$, the field dependence becomes component
dependent, and the curves may approach or even cross. This behavior reflects
the fact that the exact integrals in Eqs.~(\ref{sigma_parallel_exact})-(\ref{sigma_z_exact}) contain not only the conventional Fermi-surface weight,
but also the field-dependent constraint $|\psi(\lambda)|\leq1$, with
$\psi(\lambda)=\lambda^{2}(\lambda-\tilde{\mu})/b$. As $\tilde{\mu}$
increases, the active integration window is displaced toward larger
$\lambda$, where the Berry-curvature and orbital-magnetic-moment corrections
are weaker, reducing the relative importance of magnetic-field-induced
geometric effects.

The right panels provide a complementary view of the same physics. For small
$b$, the conductivities vary only weakly with the magnetic field, consistent
with the weak-field expansions in Eqs.~(\ref{sigma_z_main_weak})-(\ref{sigma_perp_main_weak}), where the first magnetic correction appears at
order $b^{2}$. This contrasts with the intrinsic Hall response, whose leading
term is linear in $b$. The absence of a linear contribution in the
Fermi-surface conductivities follows from the angular symmetry of the two
roots selected by the Fermi-surface constraint: odd-in-field terms cancel
after angular integration, leaving the first anisotropic corrections at
quadratic order.

At intermediate fields, clear deviations from the perturbative behavior
develop. In particular, for the larger chemical potential $\tilde{\mu}=1.3$,
the conductivities display a shallow minimum as a function of $b$, most
clearly visible in $\sigma_{\perp}^{\mathrm{FS}}$ and
$\sigma_{z}^{\mathrm{FS}}$. This nonmonotonicity results from the competition
between two effects. On the one hand, increasing the magnetic field enlarges
and deforms the region of phase space selected by the condition
$|\psi(\lambda)|\leq1$. On the other hand, the same orbital-magnetic-moment
correction redistributes the Fermi-surface weight in a way that can initially
suppress the corresponding transport channel. Once the field becomes
sufficiently large, the enlargement of the active phase space dominates and
the conductivity increases again.

The comparison between Figs.~\ref{Sigma_parallel}, \ref{Sigma_perp}, and
\ref{Sigma_z} also makes explicit the anisotropy generated by the magnetic
field. In the weak-field limit, Eq.~(\ref{sigma_FS_isotropic}) shows that
$\sigma_{\parallel}^{\mathrm{FS}}$ and $\sigma_{\perp}^{\mathrm{FS}}$ share
the same leading contribution, whereas anisotropy enters only through the
$b^{2}$ terms in Eqs.~(\ref{sigma_parallel_main_weak}) and
(\ref{sigma_perp_main_weak}). The full nonperturbative curves show that this
anisotropy becomes more pronounced at finite fields, with the perpendicular
and out-of-plane components displaying a stronger nonlinear dependence on
$b$. Thus, the magnetic field not only changes the magnitude of the
Fermi-surface response, but also reshapes its tensor structure by selecting
different transport channels relative to the direction of $\mathbf{B}$ and to
the plane of the nodal ring.

Overall, Figs.~\ref{Sigma_parallel}-\ref{Sigma_z} demonstrate that the
Fermi-surface contribution remains Drude-like in its leading dependence on
carrier density, but acquires strongly nontrivial magnetic-field corrections
once the orbital magnetic moment is retained nonperturbatively. These
corrections are quadratic in the weak-field regime, become nonlinear at
intermediate fields, and are most visible when the chemical potential lies
close to the gap, where the Fermi surface probes the geometrically active
region surrounding the nodal ring.

The field-induced anisotropy discussed above has a direct experimental
manifestation in the form of a planar Hall effect. Since the conductivity
tensor is no longer isotropic in the plane perpendicular to the nodal ring,
its components depend on the angle $\alpha$ between the applied magnetic
field and the electric field. In particular, the off-diagonal conductivity
takes the simple form
\begin{align}
\sigma_{xy}^{\mathrm{FS}}
=
\frac{1}{2} ( \, 
\sigma_{\parallel}^{\mathrm{FS}}
-
\sigma_{\perp}^{\mathrm{FS}} \, ) \,
\sin(2\alpha),
\label{sigma_xy_planar}
\end{align}
which immediately shows that the planar Hall response is entirely governed by
the magnetic-field-induced anisotropy. In the absence of a magnetic field,
$\sigma_{\parallel}^{\mathrm{FS}}=\sigma_{\perp}^{\mathrm{FS}}$, and the
planar Hall conductivity vanishes identically. The maximum Hall response is
therefore given by
\begin{align}
\max |\sigma_{xy}^{\mathrm{FS}}|
=
\frac12
\left|
\sigma_{\parallel}^{\mathrm{FS}}
-
\sigma_{\perp}^{\mathrm{FS}}
\right|,
\label{sigma_xy_max}
\end{align}
which provides a direct measure of the transport anisotropy generated by the
orbital magnetic moment.

\begin{figure}
    \centering
    \includegraphics[width=0.47\linewidth]{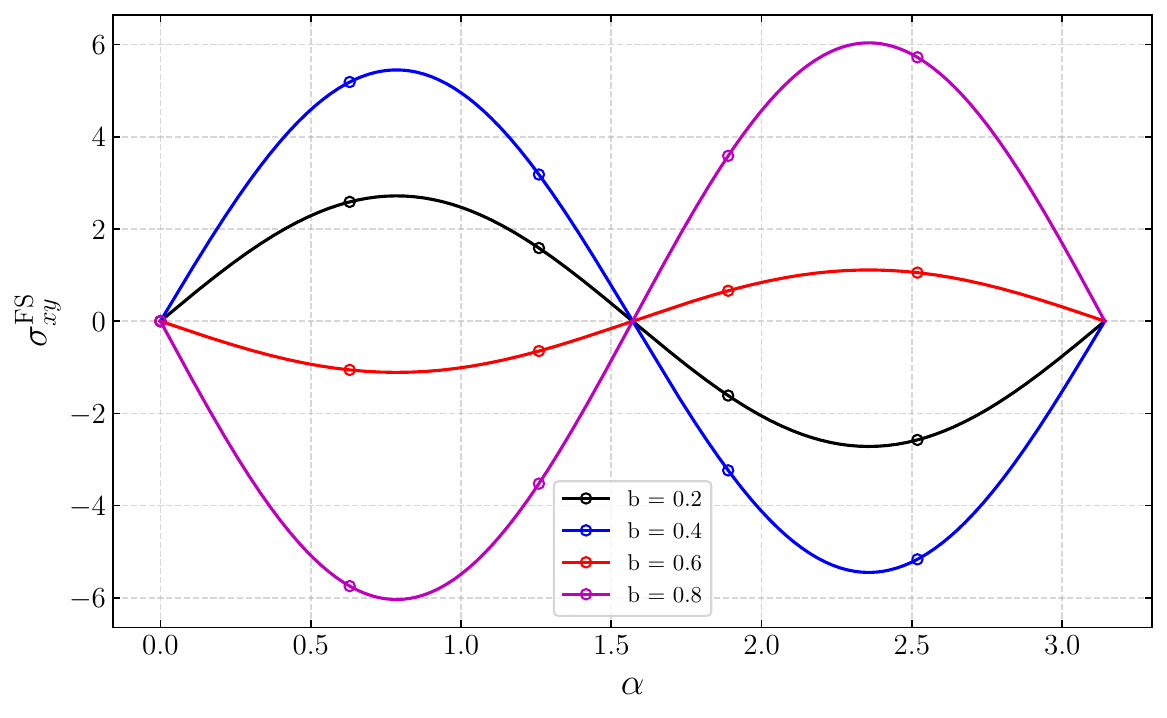}
    \includegraphics[width=0.47\linewidth]{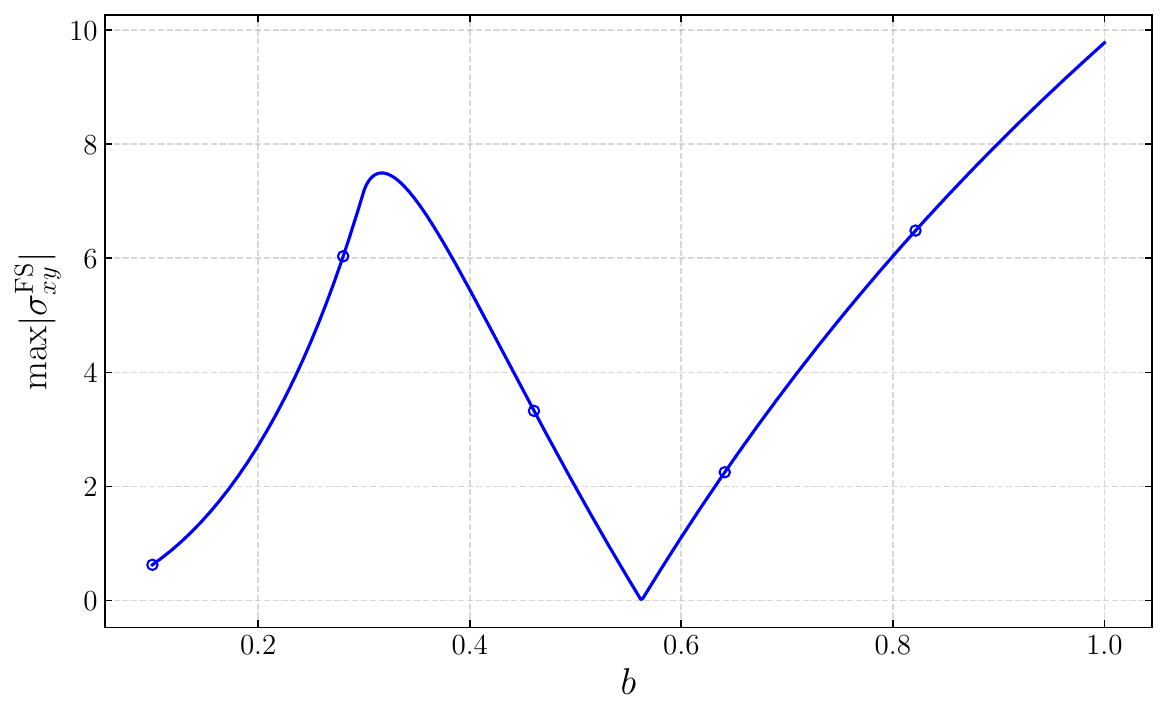}
    \caption{
Planar Hall conductivity arising from the magnetic-field-induced anisotropy of the Fermi-surface transport. Left panel: angular dependence of the planar Hall conductivity $\sigma_{xy}^{\mathrm{FS}}$, given by Eq.~(\ref{sigma_xy_planar}), for $\tilde{\mu}=1.3$ and four representative magnetic fields, $b=0.2$, $0.4$, $0.6$, and $0.8$. The response follows the characteristic $\sin(2\alpha)$ dependence of Eq.~(\ref{sigma_xy_planar}), vanishing when the magnetic field is parallel or perpendicular to the electric field ($\alpha=0,\pi/2,\pi$) and reaching its extrema at $\alpha=\pi/4$ and $3\pi/4$. Right panel: maximum planar Hall conductivity,
$\max|\sigma_{xy}^{\mathrm{FS}}|=\frac12|\sigma_{\parallel}^{\mathrm{FS}}-\sigma_{\perp}^{\mathrm{FS}}|$,
as a function of the magnetic field. The finite response directly reflects the magnetic-field-induced transport anisotropy and exhibits a strongly nonlinear evolution beyond the weak-field regime.
}    \label{Sigma_xy}
\end{figure}

The corresponding numerical results are presented in
Fig.~\ref{Sigma_xy}. The left panel shows the angular dependence of the
planar Hall conductivity for several representative magnetic fields. As
expected from Eq.~(\ref{sigma_xy_planar}), all curves follow the universal
$\sin(2\alpha)$ angular dependence, vanishing for
$\alpha=0$, $\pi/2$, and $\pi$, where the magnetic field is either parallel
or perpendicular to the applied electric field, and reaching their extrema
at $\alpha=\pi/4$ and $3\pi/4$. While the angular dependence is fixed by the
tensorial structure of the conductivity, its amplitude increases
substantially with the magnetic field, reflecting the progressive separation
between the longitudinal and transverse transport channels.

The right panel displays the maximum planar Hall conductivity,
Eq.~(\ref{sigma_xy_max}), as a function of the magnetic field. In the
weak-field regime, the response increases quadratically with $b$, in
agreement with the analytical expansions of
Eqs.~(\ref{sigma_parallel_main_weak}) and
(\ref{sigma_perp_main_weak}), where the anisotropy first appears at order
$b^{2}$. As the magnetic field increases, however, the response deviates
significantly from this perturbative behavior and eventually reaches a broad
maximum before decreasing at larger fields. This nonmonotonic evolution
originates from the same mechanism discussed throughout this work: the
orbital magnetic moment continuously reconstructs the geometrically active
Fermi surface by modifying both the quasiparticle energy and the available
phase space. Initially, increasing the magnetic field enhances the
distinction between the longitudinal and transverse transport channels,
thereby strengthening the planar Hall effect. At larger fields, however, the
field-induced redistribution of the active electronic states becomes
progressively less favorable for transport, reducing the conductivity
anisotropy and consequently the planar Hall response.

These results demonstrate that the planar Hall effect provides a direct
experimental probe of the magnetic-field-induced reconstruction of the Fermi
surface predicted by our theory. Since its magnitude is controlled entirely
by the difference
$\sigma_{\parallel}^{\mathrm{FS}}-\sigma_{\perp}^{\mathrm{FS}}$, the planar
Hall conductivity offers a particularly sensitive signature of the
nonperturbative geometric transport induced by the orbital magnetic moment in
gapped nodal-line semimetals.

\section{Conclusions} \label{conclusion}

In this work we have developed a fully nonperturbative semiclassical theory of magnetotransport in gapped nodal-line semimetals, retaining the complete magnetic-field dependence of the Berry-curvature corrections, orbital magnetic moment, generalized velocity, and phase-space factor. Unlike conventional weak-field approaches, the present formulation remains valid to all orders in the magnetic field within the semiclassical regime and provides exact nonperturbative integral representations for both intrinsic and dissipative transport coefficients.

A central result of this work is the identification of two qualitatively distinct geometric transport mechanisms. The first is an intrinsic Hall response arising from the anomalous velocity. Although the Berry curvature forms closed loops around the nodal ring and therefore produces no Hall current in the absence of a magnetic field, we show that the orbital magnetic moment modifies the occupation of electronic states in an angle-dependent manner, thereby reconstructing the geometrically active regions of the Fermi surface. This mechanism breaks the exact azimuthal cancellation characteristic of nodal-line semimetals and activates a finite Hall current whose magnitude is entirely controlled by the magnetic field. The resulting Hall conductivity exhibits a strongly nonlinear dependence on both the magnetic field and the chemical potential, providing a clear manifestation of magnetic-field-activated geometric transport.

The second mechanism originates from the nonequilibrium correction to the distribution function and describes dissipative transport at the Fermi surface. We derived exact nonperturbative expressions for the three independent conductivity components and showed that they recover the expected Drude behavior in the weak-field limit, while exhibiting quadratic magnetic-field corrections together with a progressively stronger transport anisotropy at finite fields. The numerical results further reveal that the conductivity evolves nonlinearly with magnetic field and displays shallow minima at intermediate fields, reflecting the competition between the field-induced deformation of the Fermi surface and the redistribution of geometrically active states produced by the orbital magnetic moment. An immediate consequence of this anisotropy is the emergence of a finite planar Hall conductivity, whose amplitude is directly proportional to the difference between the longitudinal and transverse conductivities. We showed that this response is quadratic in the weak-field regime, becomes strongly nonperturbative at intermediate magnetic fields, and therefore provides a direct experimental signature of the magnetic-field-induced reconstruction of the active Fermi surface.

An important outcome of our analysis is that both intrinsic and dissipative transport are governed by the same underlying geometric mechanism. In both cases, the orbital magnetic moment continuously reconstructs the region of momentum space that contributes to transport by modifying the field-corrected energy and the corresponding phase-space constraint. This interpretation provides a unified picture of magnetic-field-activated transport in nodal-line semimetals, in which the magnetic field not only modifies the quasiparticle dynamics but also determines which regions of the Fermi surface remain electronically active. From this perspective, the intrinsic Hall response, the anisotropic Fermi-surface conductivity, and the planar Hall effect should be viewed as complementary manifestations of the same geometric reconstruction induced by the orbital magnetic moment.

The present work also opens several directions for future investigation. The formalism developed here can be naturally generalized to more complex nodal-line materials containing multiple nodal rings or linked nodal structures, as well as to systems with broken inversion or time-reversal symmetry, where additional geometric transport phenomena are expected. Another promising direction is the extension of the present framework to nonlinear thermoelectric and magneto-optical responses, where the interplay between Berry curvature, orbital magnetic moment, and magnetic-field-induced Fermi-surface reconstruction may produce qualitatively new transport signatures.

Overall, our results demonstrate that the orbital magnetic moment is not merely a quantitative correction to semiclassical dynamics, but rather a fundamental mechanism that reconstructs the geometrically active Fermi surface and governs both intrinsic and dissipative magnetotransport in nodal-line semimetals beyond the conventional weak-field regime. The predicted nonlinear intrinsic Hall response, field-induced transport anisotropy, and planar Hall effect constitute experimentally accessible signatures of this geometric reconstruction, providing clear fingerprints of orbital-magnetic-moment physics in gapped nodal-line semimetals. We expect that the nonperturbative framework developed here will provide a useful basis for understanding geometric transport phenomena in a broad class of topological quantum materials.

\

\acknowledgements{L.M.O. and L.E.S.-A. were supported by SECIHTI fellowships No. 834773 and No. 4066288, respectively. A.M.-R. acknowledges financial support by UNAM-PAPIIT project No. IG100224, UNAM-PAPIME project No. PE109226, by SECIHTI project No. CBF-2025-I-1862 and by the Marcos Moshinsky Foundation.}
 
\

\appendix

\section{Evaluation of the angular integral for the intrinsic Hall contribution}
\label{Angular_Integral}

In this Section we evaluate the angular integral
\begin{align}
    \mathbf{Q} = \int _{0} ^{2 \pi } d \phi  \; \hat{\mathbf{e}} _{\phi} \, \Theta \left( \mu -  \lambda  -  \frac{e mv ^{2} \hbar B }{2 \lambda ^{2}  } \;  \hat{\mathbf{B}} \cdot  \hat{\mathbf{e}} _{\phi} \right) , \label{Ang_Int_App}
\end{align}
appearing in Eq. (\ref{J_intrinsic2}). The only external vector in the problem is the magnetic field $\mathbf{B}$. Since the integrand depends on $\hat{\mathbf{e}}_{\phi}$ only through the projection $\hat{\mathbf{B}} \cdot \hat{\mathbf{e}}_{\phi}$, rotational symmetry implies that $\mathbf{Q}$ must be collinear with $\mathbf{B}$, while any perpendicular component averages to zero. Therefore, the integral can be written in the general form
\begin{align}
    \mathbf{Q} = Q(B) \, \hat{\mathbf{B}}, \label{Ansatz_Ang_Int}
\end{align}
where $Q(B)$ is a scalar function of $B$. Taking the scalar product $\hat{\mathbf{B}} \cdot \mathbf{Q}$ and using Eq. (\ref{Ang_Int_App}) we find
\begin{align}
    Q(B) = \int _{0} ^{2 \pi } d \phi  \;  \hat{\mathbf{B}} \cdot  \hat{\mathbf{e}} _{\phi} \, \Theta \left( \mu -  \lambda  -  \frac{e mv ^{2} \hbar B }{2 \lambda ^{2}  } \;  \hat{\mathbf{B}} \cdot  \hat{\mathbf{e}} _{\phi} \right) . \label{Q_funct}
\end{align}
To evaluate this integral we introduce
\begin{align}
    a (\lambda ) = \mu - \lambda   , \qquad b (\lambda ) =  \frac{e mv ^{2} \hbar B }{2 \lambda ^{2}}  . 
\end{align}
Choosing coordinates such that $\mathbf{B}$ points along the $y$ direction, we write
\begin{align}
    Q(B) =  \int_{0}^{2\pi} d \phi \, \cos \phi \;
    \Theta \left[ a ( \lambda ) - b ( \lambda ) \cos \phi \right].
\end{align}
Assuming $b ( \lambda ) > 0$, the condition imposed by the step function is
\begin{align}
    \cos \phi < \frac{a ( \lambda ) }{ b ( \lambda )}
    \equiv p ( \lambda ) .
\end{align}
We then distinguish three cases.
\begin{itemize}
    \item If $p(\lambda)\ge 1$, the inequality is satisfied for all $\phi$ (up to a set of measure zero), and therefore
    \begin{align}
        Q(B) =  \int _{0} ^{2\pi} d \phi \, \cos \phi = 0 .
    \end{align}

\item  If $p(\lambda)\le -1$, no angle satisfies the inequality (again up to a set of measure zero), so that $Q(B) =0$.

\item If $|p(\lambda)|<1$, we define
    \begin{align}
        \phi_{0}(\lambda)=\arccos\!\bigl[p(\lambda)\bigr],
    \end{align}
and the allowed angular region is
    \begin{align}
        \phi\in[\phi_{0}(\lambda),\,2\pi-\phi_{0}(\lambda)].
    \end{align}
Hence,
\begin{align}
    Q(B) = - 2 \sin \phi_{0}(\lambda) = - 2  \sqrt{1-p^{2}(\lambda)} .
\end{align}

\end{itemize}
Substituting this result in Eq. (\ref{Ansatz_Ang_Int}) we get the required angular integral (\ref{Ang_Int_App}) which is used in the main text to obtain Eq. (\ref{J_intrinsic3}).

\section{Evaluation of the angular integral for the Fermi-surface contribution} \label{App_FS_current}

In this Appendix we evaluate the angular integral entering the Fermi-surface transport current $\mathbf{J} ^{\mathrm{FS}}$ and reduce the conductivity tensor to a one-dimensional radial integral. From Eq. (\ref{J_FS2}) we read the conductivity tensor
\begin{align}
\sigma _{ij} ^{\mathrm{FS}} = e ^{2} \tau \int \frac{d ^{3} \mathbf{k}}{( 2 \pi ) ^{3} } \; D _{\mathbf{k}} \, \mathcal{V} _{i} \mathcal{V} _{j} \, \delta \left( \mu - \lambda _{\mathbf{k}} - \frac{e m v ^{2} \hbar }{ 2 \lambda _{\mathbf{k}} ^{2} } \,\mathbf{B} \cdot \hat{\mathbf{e}} _{\phi} \right) . \label{sigma_FS_app_start}
\end{align}
We now consider a general in-plane magnetic field $\mathbf{B} = B \,\hat{\mathbf{B}}$ with $\hat{\mathbf{B}} = \cos \phi _{B} \, \hat{\mathbf{e}} _{x} + \sin \phi _{B} \, \hat{\mathbf{e}} _{y}$. Hence
\begin{align}
    \hat{\mathbf{B}} \cdot \hat{\mathbf{k}} _{\parallel} = \cos (\phi   - \phi _{B}  ) , \qquad  \hat{\mathbf{B}} \cdot \hat{\mathbf{e}} _{\phi} = - \sin (\phi   - \phi _{B}  ) . 
\end{align}
Next we introduce the decomposition of the generalized velocity (\ref{Generalized_velocity_NLSM}) as
\begin{align}
\boldsymbol{\mathcal V}_{n \mathbf{k}} =  A (\lambda _{\mathbf{k}} , \phi ) \, \mathbf{u} - \frac{C (\lambda _{\mathbf{k}} , \phi )}{k _{\rho }}  \, \hat{\mathbf e}_{\phi} + \frac{G (\lambda _{\mathbf{k}} , \phi )}{k _{\rho }} \, \hat{\mathbf{B}} ,
\end{align}
where we defined
\begin{align}
    \mathbf{u} = q _{\rho}\, (\cos \phi \, \hat{\mathbf{e}} _{x} + \sin \phi \, \hat{\mathbf{e}} _{y}) + k _{z} \hat{\mathbf{e}} _{z} , \qquad A (\lambda _{\mathbf{k}} , \phi ) = \frac{\hbar v ^{2} }{\lambda _{\mathbf{k}}} + e B m \hbar ^{2} v ^{4} \; \frac{ \sin (\phi   - \phi _{B}  ) }{\lambda _{\mathbf{k}} ^{4} } \\[5pt] C (\lambda _{\mathbf{k}} , \phi )  = \frac{e B m v ^2}{2} \frac{ \cos (\phi   - \phi _{B}  ) }{ \lambda _{\mathbf{k}} ^{2} } ,  \qquad  G (\lambda _{\mathbf{k}} , \phi ) =  \frac{e^2 B ^{2} m^2 v^4 \hbar}{4} \frac{   \cos (\phi   - \phi _{B}  ) }{ \lambda _{\mathbf{k}} ^{5} } 
\end{align}
and the phase-space factor becomes
\begin{align}
    D _{ \mathbf{k}} \equiv D (\lambda  _{\mathbf{k}} , \phi ) = \left[ 1 + \frac{e}{\hbar} \frac{ m v ^{2} \hbar ^{2} B}{2 \lambda ^{3} _{\mathbf{k}} } \,  \sin (\phi   - \phi _{B}  )  \right] ^{-1} .
\end{align}
Using the toroidal coordinates defined by Eq. (\ref{toroidal_coordinates}) we find $d ^{3} \mathbf{k} = r (k _{0} + r \cos \theta ) \, dr \, d \theta\, d \phi$, $\lambda _{\mathbf{k}} = \sqrt{m^2 + (\hbar v r) ^{2} } \equiv \lambda (r) $ and $k _{\rho} = k _{0} + r \cos \theta$. Therefore, the conductivity tensor (\ref{sigma_FS_app_start}) simplifies to 
\begin{align}
\sigma _{ij} ^{\mathrm{FS}} = \frac{e ^{2} \tau}{8 \pi ^{3} } \int _{0} ^{k _{0}} dr \int _{0} ^{2 \pi} d \theta \int _{0} ^{2 \pi} d \phi \; r (k _{0} + r \cos \theta ) \; D (\lambda , \phi )\, \left[ r A (\lambda , \phi ) \, \hat{u} _{i} - \frac{C (\lambda , \phi )}{ k _{0} + r \cos \theta }  \, ( \hat{\mathbf e}_{\phi} ) _{i} + \frac{G (\lambda , \phi )}{ k _{0} + r \cos \theta } \, \hat{B} _{i} \right]   \notag \\ \times \left[ r A (\lambda , \phi ) \, \hat{u} _{j} - \frac{C (\lambda , \phi )}{ k _{0} + r \cos \theta }  \, ( \hat{\mathbf e}_{\phi} ) _{j} + \frac{G (\lambda  , \phi )}{ k _{0} + r \cos \theta } \, \hat{B} _{j} \right] \;  \delta \left( \mu - \lambda + \frac{e m B v ^{2} \hbar }{ 2 \lambda ^{2} } \, \sin (\phi   - \phi _{B}  )  \right) . \label{sigma_FS_app_start_2}
\end{align}
where $\hat{\mathbf{u}} = \cos \theta  \, (\cos \phi \, \hat{\mathbf{e}} _{x} + \sin \phi \, \hat{\mathbf{e}} _{y}) +  \sin \theta  \, \hat{\mathbf{e}} _{z}$. 

Performing the integration over $\theta$, all terms linear in $\mathbf{u}$ vanish by symmetry. Thus we obtain
\begin{align}
\sigma _{ij} ^{\mathrm{FS}} = \frac{e ^{2} \tau}{8 \pi ^{3} } \int _{0} ^{k _{0}} dr \int _{0} ^{2 \pi} d \theta \int _{0} ^{2 \pi} d \phi \; r \, (k _{0} + r \cos \theta ) \; D (\lambda , \phi )\, \Bigg\{\ \!\! r ^{2} A ^{2} (\lambda , \phi ) \, \hat{u} _{i} \hat{u} _{j} + \frac{1}{ ( k _{0} + r \cos \theta ) ^{2} }  \left[   C (\lambda , \phi ) \, ( \hat{\mathbf e}_{\phi} ) _{i} - G (\lambda , \phi ) \, B _{i} \right] \notag \\ \times \left[  C (\lambda , \phi ) \, ( \hat{\mathbf e}_{\phi} ) _{j} - G (\lambda , \phi ) \, B _{j} \right] \Bigg\}\    \delta \left( \mu - \lambda + \frac{e m v ^{2} \hbar }{ 2 \lambda ^{2} } \, \sin (\phi   - \phi _{B}  )  \right) . \label{sigma_FS_app_start_3}
\end{align}
The nonzero contributions when integrating over the transverse angle are 
\begin{align}
    \int _{0} ^{2 \pi} d \theta \; (k _{0} + r \cos \theta ) \, \hat{u} _{i} \hat{u} _{j} = \pi k _{0} \, [ (\hat{k}_{\parallel}) _{i} \, (\hat{k}_{\parallel}) _{j} + \delta _{iz} \delta _{jz} ] , \qquad \int _{0} ^{2 \pi} \frac{d \theta}{k _{0} + r \cos \theta}  = \frac{2 \pi }{\sqrt{ k _{0} ^{2} - r ^{2} }} . 
\end{align}
Inserting these results back into Eq. (\ref{sigma_FS_app_start_3}) and changing the radial variable we get
\begin{align}
\sigma _{ij} ^{\mathrm{FS}} &= \frac{e ^{2} k _{0}  }{4 \pi h  } \frac{\tau m }{\hbar}  \, \int _{ 1 } ^{   \sqrt{1 + \Delta ^{2} } }  d \lambda \int _{0} ^{2 \pi} d \phi \; \lambda  \; \bar{D} (\lambda , \phi ) \, \Bigg\{\ \!\! \frac{ \lambda ^{2} - 1}{ \lambda ^{2} } \,  \bar{A} ^{2} ( \lambda , \phi )  \, [ (\hat{k}_{\parallel}) _{i} \, (\hat{k}_{\parallel}) _{j}  + \delta _{iz} \delta _{jz} ] + \frac{2b ^{2} / \Delta }{\sqrt{ 1 + \Delta ^{2} - \lambda ^{2} }}  \notag \\ & \hspace{3cm} \times   \bar{C} _{i} (\lambda , \phi  )  \; \bar{C} _{j} (\lambda , \phi ) \; \frac{ \cos ^{2} (\phi - \phi _{B}  ) }{ \lambda ^{4} }   \Bigg\}\   \delta \left( \tilde{\mu} -  \lambda   +   b \frac{\sin (\phi   - \phi _{B}  ) }{ \lambda ^{2} }   \right) , \label{sigma_FS_app_start_5}
\end{align}
where $\Delta = \hbar v k _{0} / m $, $b = B / B _{\mathrm{eff}}$ and
\begin{align}
    \bar{D} (\lambda , \phi ) = \left[ 1 +  b   \,  \frac{ \sin (\phi   - \phi _{B}  ) }{\lambda ^{3}} \right] ^{-1}  , \qquad  \bar{A} ( \lambda , \phi ) = 1 + 2b \frac{\sin (\phi - \phi _{B})}{\lambda ^{3}} , \qquad \bar{\mathbf{C}} (\lambda , \phi  ) = \hat{\mathbf{e}}_{\phi} - \frac{b \, }{\lambda ^{3}} \, \hat{\mathbf{B}} . 
\end{align}
To evaluate the integral in Eq. (\ref{sigma_FS_app_start_5}) we we express the Dirac delta function in terms of the roots of the argument. To this end, we have to employ the formula
\begin{align}
    \delta (g(x)) = \sum _{i} \frac{ \delta ( x-x_i ) }{\vert g'(x _i) \vert } , \label{dirac_delta_property}
\end{align}
for an arbitrary (continuously differentiable) function $g(x)$ with its roots $x_i$ defined by $g(x_i)=0$. In the present case we take $g(\phi) = \tilde{\mu} - \lambda + b \frac{\sin (\phi - \phi _{B})}{\lambda ^{2}}$.  The two roots $\phi_r$ (with $r = \pm$), determined from $g(\phi_r)=0$, are given by
\begin{align}
\phi_{+} &= \phi_B + \arcsin \psi (\lambda), \qquad \quad \phi_{-} = \phi_B + \pi - \arcsin \psi (\lambda),
\end{align}
where
\begin{align}
\psi (\lambda) \equiv \frac{\lambda ^{2}(\lambda - \tilde{\mu})}{b} \le 1 .
\end{align}
Therefore, using the identity (\ref{dirac_delta_property}) we obtain
\begin{align}
\delta \left( \tilde{\mu} -  \lambda   +   b \frac{\sin (\phi   - \phi _{B}  ) }{ \lambda ^{2} }   \right) = \frac{\lambda ^{2}}{b} \frac{ \delta(\phi - \phi_+) + \delta(\phi-\phi_-)}{ \sqrt{1 - \psi ^{2} (\lambda) } } .
\end{align}
Using this result one can further evaluate the integral with respect to $\phi$ in Eq. (\ref{sigma_FS_app_start_5}). Performing the sum over the two roots $\phi _r$ with $r=\pm$, Eq. (\ref{sigma_FS_app_start_5}) becomes
\begin{align}
\sigma _{ij} ^{\mathrm{FS}} &= \frac{e ^{2} k _{0}}{4 \pi h} \frac{\tau m }{\hbar} \int_{1}^{\sqrt{1+\Delta^2}} d\lambda \; \frac{\lambda ^{4}}{b (2\lambda-\tilde\mu)\sqrt{1-\psi^2 }} \Bigg\{ \frac{\lambda^2-1}{\lambda^4}(3\lambda-2\tilde\mu)^2 \Big[ (1-\psi^2)\, ( \hat B_i \hat B_j +   \delta_{iz}\delta_{jz} ) + \psi^2\,  \delta_{ij}   \Big] \notag\\ &\hspace{0.8cm} + \frac{2b^2/\Delta}{\sqrt{1+\Delta^2-\lambda^2}} \frac{1-\psi^2}{\lambda^4} \Bigg[ (1-\psi^2)\,(\hat B_{\perp})_i(\hat B_{\perp})_j + \left(\psi^2+\frac{b^2}{\lambda^6}\right)\hat B_i\hat B_j  - \frac{b}{\lambda^3}\psi \Big( (\hat B_{\perp})_i\hat B_j+\hat B_i(\hat B_{\perp})_j \Big) \Bigg] \Bigg\} . \label{sigma_FS_app_start_6}
\end{align}
To obtain this result we used the parametrization
\begin{align}
\phi_r = \phi_B + \theta_r,
\qquad
\sin \theta_r = \psi(\lambda),
\qquad
\cos \theta_r = r \sqrt{1-\psi^2(\lambda)},
\qquad r=\pm,
\end{align}
from which it follows that
\begin{align}
\sum_{r=\pm} \cos \theta_r = 0, \qquad \sum_{r=\pm} \sin \theta_r = 2\psi(\lambda), \qquad \sum_{r=\pm} \cos^2 \theta_r = 2\left[1-\psi^2(\lambda)\right], \\ \sum_{r=\pm} \sin^2 \theta_r = 2\psi^2(\lambda), \qquad \sum_{r=\pm} \sin\theta_r \cos\theta_r = 0.
\end{align}
In particular, using $\hat{k}_{\parallel} = \cos\phi \, \hat{\mathbf e}_x + \sin\phi \, \hat{\mathbf e}_y$  one finds
\begin{align}
\sum_{r=\pm} (\hat{k}_{\parallel})_i (\hat{k}_{\parallel})_j = 2(1-\psi^2)\,\hat B_i \hat B_j + 2 \psi^2\,(\hat B_{\perp})_i(\hat B_{\perp})_j,
\end{align}
with $\hat{\mathbf B} _{\perp} = \hat{\mathbf e} _{z} \times \hat{\mathbf B}$. Likewise, from $\bar{\mathbf C}(\lambda,\phi_r)$, we obtain
\begin{align}
\frac{1}{2}\sum_{r=\pm} \bar C_i(\lambda,\phi_r)\bar C_j(\lambda,\phi_r)
&=
(1-\psi^2)\,(\hat B_{\perp})_i(\hat B_{\perp})_j
+
\left(\psi^2+\frac{b^2}{\lambda^6}\right)\hat B_i\hat B_j
-
\frac{b}{\lambda^3}\psi
\Big(
(\hat B_{\perp})_i\hat B_j+\hat B_i(\hat B_{\perp})_j
\Big).
\end{align}
Finally, the tensor of equation (\ref{sigma_FS_app_start_5}) can be decomposed as
\begin{align}
\sigma_{ij}^{\mathrm{FS}} = \sigma_{\parallel}^{\mathrm{FS}} \,\hat B_i \hat B_j + \sigma_{\perp}^{\mathrm{FS}}\, (\hat B_{\perp})_i (\hat B_{\perp})_j + \sigma_z^{\mathrm{FS}}\,\delta_{iz}\delta_{jz} ,
\end{align}
where we introduced
\begin{align}
\sigma_{\parallel}^{\mathrm{FS}}
&=
\sigma ^{\mathrm{Hall}} \; \frac{2 \tau m }{\hbar}
\int_{1}^{\sqrt{1+\Delta^2}} d\lambda \;
\frac{\lambda ^{4}}{b (2\lambda-\tilde\mu)\sqrt{1-\psi^2 }}
\left[
\frac{\lambda^2-1}{\lambda^4}(3\lambda-2\tilde\mu)^2(1-\psi^2)
+
\frac{2b^2/\Delta}{\sqrt{1+\Delta^2-\lambda^2}}
\frac{1-\psi^2}{\lambda^4}
\left(\psi+\frac{b}{\lambda^3}\right)^2
\right],
\\
\sigma_{\perp}^{\mathrm{FS}}
&=
\sigma ^{\mathrm{Hall}} \; \frac{2 \tau m }{\hbar}
\int_{1}^{\sqrt{1+\Delta^2}} d\lambda \;
\frac{\lambda ^{4}}{b (2\lambda-\tilde\mu)\sqrt{1-\psi^2 }}
\left[
\frac{\lambda^2-1}{\lambda^4}(3\lambda-2\tilde\mu)^2\psi^2
+
\frac{2b^2/\Delta}{\sqrt{1+\Delta^2-\lambda^2}}
\frac{(1-\psi^2)^2}{\lambda^4}
\right],
\\
\sigma_{z}^{\mathrm{FS}}
&=
\sigma ^{\mathrm{Hall}} \; \frac{2 \tau m }{\hbar}
\int_{1}^{\sqrt{1+\Delta^2}} d\lambda \;
\frac{\lambda ^{4}}{b (2\lambda-\tilde\mu)\sqrt{1-\psi^2 }}
\left[
\frac{\lambda^2-1}{\lambda^4}(3\lambda-2\tilde\mu)^2
\right].
\end{align}
These expressions define the exact radial integrals used in the numerical evaluation of the Fermi-surface transport coefficients.

For a better understanding of the structure of these exact expressions, it is useful to analyze their weak-field limit. In the present dimensionless notation, this regime corresponds to $b \ll 1$, together with the condition that the Fermi level lies well inside the integration interval, $1<\tilde\mu<\sqrt{1+\Delta^2}$, so that the narrow region selected by the constraint $|\psi(\lambda)|\le 1$ remains away from the endpoints. Since $\psi(\lambda)= \lambda^2(\lambda-\tilde\mu) / b$, the condition $|\psi(\lambda)|\le 1$ implies that, for $b\ll 1$, the radial integral is restricted to a small neighborhood around $\lambda=\tilde\mu$, with characteristic width
\begin{align}
|\lambda-\tilde\mu|\sim \frac{b}{\tilde\mu^2}.
\label{eq:lambda_width}
\end{align}
It is then convenient to introduce the variable
\begin{align}
u=\psi(\lambda)=\frac{\lambda^2(\lambda-\tilde\mu)}{b},
\label{eq:u_variable}
\end{align}
for which
\begin{align}
\frac{du}{d\lambda}
=
\frac{3\lambda^2-2\tilde\mu\lambda}{b}
=
\frac{\lambda(3\lambda-2\tilde\mu)}{b},
\qquad
d\lambda=\frac{b}{\lambda(3\lambda-2\tilde\mu)}\,du.
\label{eq:du_dlambda}
\end{align}
Then for the $z$-component the conductivity becomes
\begin{align}
\sigma_{z}^{\mathrm{FS}}
=
\sigma ^{\mathrm{Hall}} \; \frac{2 \tau m }{\hbar}
\int_{-1}^{1} du\;
\frac{(\lambda^2-1)(3\lambda-2\tilde\mu)}
{\lambda(2\lambda-\tilde\mu)\sqrt{1-u^2}},
\label{eq:sigma_z_u}
\end{align}
where $\lambda=\lambda(u)$ is determined implicitly by $u=\lambda^2(\lambda-\tilde\mu)/b$. In the weak-field regime, the inverse relation around $\lambda = \tilde\mu$ reads
\begin{align}
\lambda
=
\tilde\mu+\frac{b}{\tilde\mu^2}u-\frac{2b^2}{\tilde\mu^5}u^2+ \mathcal{O} (b^3).
\label{eq:lambda_of_u_expansion}
\end{align}
Expanding the full integrand, one finds
\begin{align}
\frac{(\lambda^2-1)(3\lambda-2\tilde\mu)}
{\lambda(2\lambda-\tilde\mu)}
=
\frac{\tilde\mu^2-1}{\tilde\mu}
+
\frac{2b}{\tilde\mu^2}u
+
\frac{2-5\tilde\mu^2}{\tilde\mu^7}b^2u^2
+
\mathcal{O} (b^3).
\label{eq:common_expansion}
\end{align}
Therefore,
\begin{align}
\sigma_{z}^{\mathrm{FS}}
=
\sigma ^{\mathrm{Hall}} \; \frac{2 \tau m }{\hbar}
\int_{-1}^{1} du\;
\frac{1}{\sqrt{1-u^2}}
\left[
\frac{\tilde\mu^2-1}{\tilde\mu}
+
\frac{2b}{\tilde\mu^2}u
+
\frac{2-5\tilde\mu^2}{\tilde\mu^7}b^2u^2
\right]
+
\mathcal{O} (b^3) .
\label{eq:sigma_z_expanded}
\end{align}
Using
\begin{align}
\int_{-1}^{1}\frac{du}{\sqrt{1-u^2}}=\pi,
\qquad
\int_{-1}^{1}\frac{u\,du}{\sqrt{1-u^2}}=0,
\qquad
\int_{-1}^{1}\frac{u^2\,du}{\sqrt{1-u^2}}=\frac{\pi}{2},
\label{eq:basic_u_integrals}
\end{align}
we finally obtain
\begin{align}
\sigma_{z}^{\mathrm{FS}}
=
\sigma ^{\mathrm{Hall}} \; \frac{2 \pi \tau m }{\hbar}
\left[
\frac{\tilde\mu^2-1}{\tilde\mu}
+
\frac{b^2}{2}\frac{2-5\tilde\mu^2}{\tilde\mu^7}
\right]
+
\mathcal{O} (b^3) .
\label{eq:sigma_z_weak_field}
\end{align}

The same procedure can be applied to the longitudinal and transverse Fermi-surface conductivities introduced in Eqs.~(\ref{sigma_parallel_exact}) and (\ref{sigma_perp_exact}). After changing variables according to Eqs.~(\ref{eq:u_variable})-(\ref{eq:du_dlambda}), one finds
\begin{align}
\sigma_{\parallel}^{\mathrm{FS}}
&=
\sigma ^{\mathrm{Hall}} \; \frac{2 \tau m }{\hbar}
\int_{-1}^{1} du\;
\Bigg\{
\sqrt{1-u^2}\;
\frac{(\lambda^2-1)(3\lambda-2\tilde\mu)}
{\lambda(2\lambda-\tilde\mu)}
+
\frac{2b^2/\Delta}{\sqrt{1+\Delta^2-\lambda^2}}\,
\frac{\sqrt{1-u^2}}{\lambda(3\lambda-2\tilde\mu)(2\lambda-\tilde\mu)}
\left(u+\frac{b}{\lambda^3}\right)^2
\Bigg\},
\label{eq:sigma_parallel_u}
\\[6pt]
\sigma_{\perp}^{\mathrm{FS}}
&=
\sigma ^{\mathrm{Hall}} \; \frac{2 \tau m }{\hbar}
\int_{-1}^{1} du\;
\Bigg\{
\frac{u^2}{\sqrt{1-u^2}}\;
\frac{(\lambda^2-1)(3\lambda-2\tilde\mu)}
{\lambda(2\lambda-\tilde\mu)}
+
\frac{2b^2/\Delta}{\sqrt{1+\Delta^2-\lambda^2}}\,
\frac{(1-u^2)^{3/2}}{\lambda(3\lambda-2\tilde\mu)(2\lambda-\tilde\mu)}
\Bigg\},
\label{eq:sigma_perp_u}
\end{align}
where again $\lambda=\lambda(u)$ is given perturbatively by Eq.~(\ref{eq:lambda_of_u_expansion}).

Using the expansion in Eq.~(\ref{eq:common_expansion}), the first terms in Eqs.~(\ref{eq:sigma_parallel_u}) and (\ref{eq:sigma_perp_u}) are obtained directly. For the terms already proportional to $b^2$, it is sufficient to evaluate the smooth $\lambda$-dependent factors at $\lambda=\tilde\mu$, since any further correction would only contribute at order $b^3$ or higher. In this way, one finds
\begin{align}
\sigma_{\parallel}^{\mathrm{FS}}
&=
\sigma ^{\mathrm{Hall}} \; \frac{2 \tau m }{\hbar}
\int_{-1}^{1} du\;
\Bigg[
\sqrt{1-u^2}
\left(
\frac{\tilde\mu^2-1}{\tilde\mu}
+
\frac{2b}{\tilde\mu^2}u
+
\frac{2-5\tilde\mu^2}{\tilde\mu^7}b^2u^2
\right) 
+
\frac{2b^2}{\Delta\,\tilde\mu^3\sqrt{1+\Delta^2-\tilde\mu^2}}\,
u^2\sqrt{1-u^2}
\Bigg]
+
\mathcal{O}(b^3),
\label{eq:sigma_parallel_expanded}
\\[6pt]
\sigma_{\perp}^{\mathrm{FS}}
&=
\sigma ^{\mathrm{Hall}} \; \frac{2 \tau m }{\hbar}
\int_{-1}^{1} du\;
\Bigg[
\frac{u^2}{\sqrt{1-u^2}}
\left(
\frac{\tilde\mu^2-1}{\tilde\mu}
+
\frac{2b}{\tilde\mu^2}u
+
\frac{2-5\tilde\mu^2}{\tilde\mu^7}b^2u^2
\right) 
+
\frac{2b^2}{\Delta\,\tilde\mu^3\sqrt{1+\Delta^2-\tilde\mu^2}}\,
(1-u^2)^{3/2}
\Bigg]
+
\mathcal{O}(b^3).
\label{eq:sigma_perp_expanded}
\end{align}
In addition to Eq.~(\ref{eq:basic_u_integrals}), we also use
\begin{align}
\int_{-1}^{1}\sqrt{1-u^2}\,du=\frac{\pi}{2},
\qquad
\int_{-1}^{1}u^2\sqrt{1-u^2}\,du=\frac{\pi}{8},
\qquad
\int_{-1}^{1}\frac{u^4\,du}{\sqrt{1-u^2}}=\frac{3\pi}{8},
\qquad
\int_{-1}^{1}(1-u^2)^{3/2}du=\frac{3\pi}{8}.
\label{eq:extra_u_integrals}
\end{align}
The odd terms again vanish after integration, and the final weak-field expansions become
\begin{align}
\sigma_{\parallel}^{\mathrm{FS}}
&=
\sigma ^{\mathrm{Hall}} \; \frac{\pi \tau m }{\hbar}
\left[
\frac{\tilde\mu^2-1}{\tilde\mu}
+
\frac{b^2}{4}\frac{2-5\tilde\mu^2}{\tilde\mu^7}
+
\frac{b^2}{2\,\Delta\,\tilde\mu^3\sqrt{1+\Delta^2-\tilde\mu^2}}
\right]
+
\mathcal{O}(b^3),
\label{eq:sigma_parallel_weak_field}
\\[6pt]
\sigma_{\perp}^{\mathrm{FS}}
&=
\sigma ^{\mathrm{Hall}} \; \frac{\pi \tau m }{\hbar}
\left[
\frac{\tilde\mu^2-1}{\tilde\mu}
+
\frac{3b^2}{4}\frac{2-5\tilde\mu^2}{\tilde\mu^7}
+
\frac{3b^2}{2\,\Delta\,\tilde\mu^3\sqrt{1+\Delta^2-\tilde\mu^2}}
\right]
+
\mathcal{O}(b^3).
\label{eq:sigma_perp_weak_field}
\end{align}

\bibliography{Refs_NLSM}
\end{document}